\documentclass[twocolumn,pra,nobalancelastpage,aps,raggedbottom]{revtex4-2}
\usepackage{amsmath,amsfonts,amssymb,graphicx,hyperref,epstopdf,xcolor,tikz,scalerel,natbib}
\usepackage{mathptmx,textcomp}
\usepackage[T1]{fontenc}
\usepackage[utf8]{inputenc}

\newcommand{\td}{\textdegree}
\newcommand{\thetap}{$\theta_p$}
\newcommand{\thetas}{$\theta_s$}
\newcommand{\thetae}{$\theta_e$}
\newcommand{\thetaq}{$\theta_{\mathbf{q}}$}
\newcommand{\tq}{$\theta_{-\mathbf{q}}$}
\newcommand{\m}{$m$}
\newcommand{\p}{\textit{p}}
\newcommand{\s}{\textit{s}}
\newcommand{\CH}{CH$_4$}
\newcommand{\NH}{NH$_3$}
\newcommand{\TDCSAV}{(TDCS)$_{av}$}
\newcommand{\Delm}{$\Delta m$}
\newcommand{\Dela}{$\Delta \alpha$}
\newcommand{\etal}{\textit{et al} }
\newcommand\inlineeqno{\stepcounter{equation}\ (\theequation)}
\definecolor{lime}{HTML}{A6CE39}

\usetikzlibrary{svg.path}
\definecolor{orcidlogocol}{HTML}{A6CE39}
\tikzset{
orcidlogo/.pic={
\fill[orcidlogocol] svg{M256,128c0,70.7-57.3,128-128,128C57.3,256,0,198.7,0,128C0,57.3,57.3,0,128,0C198.7,0,256,57.3,256,128z};
\fill[white] svg{M86.3,186.2H70.9V79.1h15.4v48.4V186.2z}
svg{M108.9,79.1h41.6c39.6,0,57,28.3,57,53.6c0,27.5-21.5,53.6-56.8,53.6h-41.8V79.1z M124.3,172.4h24.5c34.9,0,42.9-26.5,42.9-39.7c0-21.5-13.7-39.7-43.7-39.7h-23.7V172.4z}
svg{M88.7,56.8c0,5.5-4.5,10.1-10.1,10.1c-5.6,0-10.1-4.6-10.1-10.1c0-5.6,4.5-10.1,10.1-10.1C84.2,46.7,88.7,51.3,88.7,56.8z};
}
}
\newcommand\orcidicon[1]{\href{https://orcid.org/#1}{\mbox{\scalerel*{
\begin{tikzpicture}[yscale=-1,transform shape] \pic{orcidlogo}; \end{tikzpicture}
}{|}}}}  

\graphicspath{ {./Figures/} }
\DeclareMathAlphabet{\mathpzc}{OT1}{pzc}{m}{it}
\hypersetup{colorlinks,    citecolor=blue,    filecolor=black,    linkcolor=blue,    urlcolor=blue}
\begin{document}
\title{Dynamics of twisted electron impact ionization of \CH \ and \NH \ molecule}
\author{{Nikita Dhankhar$^1$ \orcidicon{0000-0001-9423-4796}}}
\author{{Neha$^1$ \orcidicon{0000-0002-2253-4797}}}


\author{{R. Choubisa$^1$ \orcidicon{0000-0003-3000-6174}} }
\email{rchoubisa@pilani.bits-pilani.ac.in}
\affiliation{$^1$ Department of Physics, Birla Institute of Technology and Science-Pilani, Pilani Campus, Pilani,  Rajasthan, 333031, India}

\begin{abstract}
Electron vortex beams (EVB, also known as twisted electron beams) possess an intrinsic orbital angular momentum (OAM) with respect to their propagation direction. This intrinsic OAM represents a new degree of freedom that provides new insights into investigating the dynamics of electron impact ionization. In this communication, we present, in the first-Born approximation (FBA), the angular profiles of the triple differential cross-section (TDCS) for the (e, 2e) process on \CH \ and \NH \ molecular targets in the coplanar asymmetric geometry. We compare the TDCS of the EVB for different values of OAM number \m \ with that of the plane wave. For a more realistic scenario, we investigate the average TDCS for macroscopic targets to explore the influence of the opening angle \thetap \ of the twisted electron beam on the TDCS. In addition, we also present the TDCS for the coherent superposition of {\it two} EVBs. The results demonstrate that the twisted (e, 2e) process retrieves the \p-type character of the molecular orbitals, which is absent in the plane wave TDCS for the given kinematics. The results for the coherent superposition of {\it two} Bessel beams show the sensitivity of TDCS towards the OAM number \m.
\end{abstract}

\maketitle

\section{Introduction}\label{sec1}
Studies on the potential applications of vortex beams (both optical and electron vortex beams) have increased over the past two decades, and thorough literature is available for both beams \cite{Torres2011, Molina2007, Bliokh2017, Lloyd2017, Hugo2018, Ivanov2022}. Electron vortex beams (EVBs, also termed as "twisted electron beams" or "Bessel beams") represent experimentally realizable, freely propagating beams carrying a well-defined orbital angular momentum (OAM)  about their propagation axis. Bliokh \etal \cite{Bliokh2007} stimulated the current research activity in electron vortex beams and their OAM content. Subsequently, Uchida and Tonomura \cite{Uchida2010} and  Verbeeck \etal \cite{Verbeeck2010} reported the generation of an EVB using a spiral phase plate and holographic technique in the transmission electron microscope, respectively.
With the current experimental techniques, researchers have developed EVBs having OAM up to 1000$\hbar$ \cite{Mafakheri2017, Tavabi2022}. Examples like, Energy-loss spectroscopy (EELS) and detection of chirality of crystals \cite{Juch2015, Juch2016b}, manipulation of nano-particles \cite{Gnanavel2012}, use of the inherent phase structure of EVB to image the biological samples \cite{Jesacher2005}, scattering studies \cite{Serbo2015, Dhankhar2020, Gong2022}  show the potential applications of the EVBs. To fully understand their applications, it is crucial to comprehend how the twisted electron beams interact with matter.

Electron-impact single ionization((e, 2e) process) is one
of the most fundamental processes in collision physics. In a coincident (e, 2e) process, an incident electron of definite energy ejects one of the bound electrons from the target upon interaction with the target, thus ionizing the target, and the outgoing electrons are detected with their energies and angles fully resolved. The triple-differential cross-section (TDCS) fully explains the ionization dynamics of a coincident (e, 2e) process. The (e, 2e) process on atomic and molecular targets has practical applications in fields like plasma physics, astronomy, atmospheric physics, radiation physics, and biology, to name a few \cite{Bartschat2016, DUNN2015, Girazian2017, Kyniene2019, Caleman2009}.

The study of electron impact ionization has made significant progress over time and is continually developing. A complete understanding of the collision dynamics for molecular targets is still challenging. The (e, 2e) study on molecules is both exciting and difficult due to the multi-center nature of the molecules and multiple scattering centers. Biologically relevant molecules like, Methane (\CH) and Ammonia (\NH) are of special interest among researchers \cite{Yavuz2014, Yavuz2016, Tachino2015, Toth2016}. Methane and ammonia are isoelectronic targets, ({\it i.e.} they have 10 electrons in the valence state), but they have different molecular structures. Thus, with the present study, we can investigate the influence of the molecular structure on the twisted electron (e, 2e) process. For these molecular targets, considerable research has been put into developing theoretical models that agree with the experimental TDCS for the plane wave. To name a few, one Coulomb wave (1CW), Brauner-Briggs-Klar (BBK) \cite{LB2009, Mir2015}, first Born and second Born using 1CW \cite{Mouawad2018,Bouchikhi2019}, molecular 3-body distorted wave (M3DW) \cite{Cnixon2011,Ali2019,Nixon2013}, distorted-wave Born approximation (DWBA) \cite{Toth2010}, Kohn variational method \cite{Chih2014}, generalized Sturmian function \cite{Castro2016, Castro2017}, multicenter distorted-wave method (MCDW) \cite{Gong2017}, BBK and its extensions \cite{Houamer2017}, 1CW with short range potential (1CWSR),  1DWSR and BBKSR \cite{Mir2020}.

There are relatively few theoretical studies of the twisted-electron-impact (e, 2e) process for atoms and molecules. The twisted electron impact ionization provides insight into new ways to explore the target structure, orientation, scattering dynamics, etc. The study by Harris \etal \cite{Harris2019} showed the angular distribution of the fully differential cross-section for the single ionization of an H atom by an incident vortex beam. Their results indicate ionization probability by the incident twisted electron beam is less likely than the plane-wave beam. Another study by the same group explains the dependence of the average over-impact parameter cross-section on the opening angle by studying the fully-differential cross-section for the excited state of the H atom \cite{Plumadore2020}. The study by Dhankhar \etal investigates the influence of the OAM number \m \ and the opening angle \thetap \ on single ionization cross-sections of molecular targets like, H$_2$ and H$_2$O and noble gas atoms by twisted electron waves \cite{Dhankhar2020, Dhankhar2022, Dhankhar2022_2}. The semi-relativistic (e, 2e) study by Mandal \etal \cite{Mandal2021} extends the investigation of twisted electron (e, 2e) processes to the relativistic regime. They studied the TDCS for charge-charge interaction and current-current interaction with the interference term for different parameters of the twisted electron beam. The theoretical investigations by Gong \etal \cite{Gong2022} use the multi-center distorted-wave method to examine TDCS for the ionization of H$_2$O molecule in the coplanar asymmetric geometry. The studies, as mentioned above, review the angular distributions of the triple differential cross-section. These studies indicate that angular and energy spectra measurement can provide insight into electron vortex ionization mechanisms, which may help advance the applications of structured electron beams. 

In the present study, we extend our previous theoretical investigation of the differential cross-section by the twisted electron beam for the H$_2$O to the \CH \ and \NH \ molecular targets. We present the theoretical estimation of the ionization cross-section by the twisted electron beam on \CH \ and \NH \ molecules in the coplanar asymmetric geometrical arrangement. The iso-electronic nature of the molecules makes this study exciting. Comparing the TDCS from the twisted electron beam (e, 2e) process with that of the plane wave may reveal important information about the effect of molecular structure on the ionization process. We investigate the twisted electron (e, 2e) process of the valence orbitals of the molecules. We develop our mathematical formalism in the framework of first-Born approximation (FBA) using the 1CW model. We describe the molecular wave functions by an expansion over the Slater-type functions centered at the heaviest nucleus (carbon (C) for \CH \ and nitrogen (N) for \NH) as proposed by Moccia \cite{Moccia1, Moccia2}. We describe the plane wave, Slater-type wave functions, Coulomb wave for the scattered electron, the molecular state of \CH \ and \NH, and the ejected electron, respectively. In this communication, we ignore the exchange effects between the incident/scattered and the bound/ejected electron since the energy of the incident or scattered electron is greater than that of the bound or ejected electron. We also ignore the post-collision interaction (PCI) between the scattered and ejected electrons. In our theoretical model, we consider frozen core approximation in which, for a multi-electron target, only one of the target electrons participates in the ionization process and is ejected in the final channel, while the other electrons remain frozen. The twisted electron beam propagates along the {\it z}-axis with the opening angle of the beam the same as the scattering angle (\thetap \ = \thetas \ =  6\td). We will see in this paper that the variations in the OAM number ({\it m}) and the opening angle $\theta_p$ lead to a significant change in the angular profiles of the TDCS.

We present the theoretical formalism of our calculation of the TDCS in Sec.\ref{sec2}. We report our results of the angular distributions of the TDCS for the outer orbitals of the atoms for different parameters of the twisted electron beam in Sec.\ref{sec3}. Finally, we conclude our paper in Sec.\ref{sec4}. Atomic units are used throughout the paper unless otherwise stated.

\section{Theory}\label{sec2}
The details of the theoretical method have been given in Ref. \cite{Dhankhar2022}; so here, we briefly outline the formalism.
Before discussing the twisted electron impact ionization of \CH \ and \NH \ molecules, it is essential to briefly describe the molecular structure and delineate the orbital labeling conventions for these targets.
According to molecular orbital theory, ammonia has three valence energy levels (\NH). The highest occupied molecular orbital is 3a$_1$, and the next highest occupied is 1e$_1$. Both these orbitals have a significant atomic \p-like character. The third energy level corresponds to  2a$_1$ molecular orbital having atomic \s-like characteristics. Similarly, for methane (\CH), the highest occupied molecular orbital is 1t$_2$ having a \p-type character, and the next highest occupied is 2a$_1$ having a \s-type character.

\subsection{"Twisted" electron ionization cross-section}

To investigate the (e, 2e) process, one must examine the triple differential cross section (TDCS), given by:
\begin{equation} \label{1}
\frac{d^{3}\sigma}{d\Omega_{e}d\Omega_{s}dE_{e}} = (2\pi)^{4}\frac{k_{e}k_{s}}{k_i}|T_{fi}|^{2},    
\end{equation}
where $\mathbf{k}_i, \mathbf{k}_s,$ and $\mathbf{k}_e$ are the momentum vectors of the incident, scattered, and ejected electrons, respectively. $dE_e$ describes the energy interval for the ejected electron, and $d\Omega_s$ and $d\Omega_e$ are the solid angle's intervals of the scattered and the ejected electron, respectively. $T$ is the matrix element for the transition of the system from initial state  $|\Psi_i\rangle$ to the final state $|\Psi_f\rangle$ via the interaction $V$ in the first Born approximation;
\begin{equation}\label{2}
T_{fi} = \langle \Psi_f|V|\Psi_i \rangle,  
\end{equation}
where 
\begin{equation}\label{3}
V = -\frac{Z}{r_0} + \sum_{i = 1}^{N}\frac{1}{|\mathbf{r}_0 - \mathbf{r}_i|}.
\end{equation}
Here, $\mathbf{r}_0$ and $\mathbf{r}_i$ are the position vectors of the incident electron and the $i^{th}$ electron, respectively, Z is the atomic number and N is the number of electrons in the target. A molecular target does not have a symmetrical charge distribution; hence we should consider the molecular target's orientation. Since all the orientations are equiprobable, we obtain the TDCS by taking an average over all the possible orientations of the molecule;
\begin{equation}\label{4}
\frac{d^3\sigma}{d\Omega_e d\Omega_s dE_e} = \frac{1}{8 \pi^2} \int \sigma^{(5)}(\alpha,\beta,\gamma)  \sin \beta d\alpha d\beta d\gamma,
\end{equation}
where, $\sigma^{(5)}(\alpha,\beta,\gamma)$ is the five-fold differential cross-section of a given molecular orbital given as;
\begin{equation}\label{5}
\begin{aligned}
\sigma^{(5)}(\alpha,\beta,\gamma) & = \frac{d^5\sigma}{d\omega d\Omega_e d\Omega_s dE_e}\\
& =  (2\pi)^4 \frac{k_ek_s}{k_i} |T_{fi}^{pw}|^2.
\end{aligned}
\end{equation}
In eqn. \ref{5}, $d\omega = \sin\beta d\alpha d\beta d\gamma$ is the solid angle element for the molecular orientation in the laboratory frame, and $\alpha,\beta$ and $\gamma$ are the Euler angles of the molecule. Interestingly, both \CH \ and \NH \ contain a heavy atom compared to the constituent hydrogen atoms; thus, the molecular orbital can be expressed as the linear combinations of the Slater-type functions centered around the heavy atom \cite{Moccia1, Moccia2}. In the frozen-core approximation, the matrix element $T_{fi}^{pw}$ for plane wave is;
\begin{equation}\label{6}
T_{fi}^{pw}(\mathbf{q}) = \frac{-2}{q^2} \langle \psi^-_{\mathbf{k}_e} | e^{i\mathbf{q}\cdot\mathbf{r}} - 1| \Phi_j(\mathbf{r}) \rangle,
\end{equation}
where $\mathbf{q} = \mathbf{k}_i - \mathbf{k}_s$ is the momentum transferred to the target, $\psi_{\mathbf{k}_e}^-(\mathbf{r})$ and $\Phi_j(\mathbf{r})$ represent the Coulomb wave-function and the molecular wave-function respectively.
The molecular orbital wave function is expressed by the linear combinations of the Slater-type functions given as;
\begin{equation}\label{7}
\Phi_j(\mathbf{r}) = \sum_{k = 1} ^{N_j}a_{jk} \phi_{n_{jk}l_{jk}m_{jk}}^{\xi_{jk}}(\mathbf{r}),
\end{equation}
where $N_j$ is the number of Slater functions used to describe the $j^{th}$ molecular orbital and $n_{jk}$,$l_{jk}$,$m_{jk}$ are the quantum numbers for the $j^{th}$ molecular orbital. $a_{jk}$ is the weight of each atomic component $\phi_{n_{jk}l_{jk}m_{jk}}^{\xi_{jk}}(\mathbf{r})$ and  $\xi_{jk}$ is a variational parameter.  $\phi_{n_{jk}l_{jk}m_{jk}}^{\xi_{jk}}(\mathbf{r})$ is  expressed as \cite{Champion2005};

\begin{equation}\label{8}
\phi_{n_{jk}l_{jk}m_{jk}}^{\xi_{jk}}(\mathbf{r}) = R^{\xi_{jk}}_{n_{jk}}(r) S_{l_{jk}m_{jk}}(\mathbf{\hat{r}}),
\end{equation}
where $R^{\xi_{jk}}_{n_{jk}}(r)$ is the radial part of each atomic orbital and given as;
\begin{equation}\label{9}
R^{\xi_{jk}}_{n_{jk}}(r) = \frac{(2\xi_{jk})^{n_{jk}+\frac{1}{2}}}{\sqrt{2n_{jk}!}}r^{n_{jk}-1}e^{-\xi_{jk}r},
\end{equation}
and $S_{l_{jk}m_{jk}}(\mathbf{\hat{r}})$ is the real spherical harmonics expressed as,

for $m_{jk} \neq 0$: 
\begin{equation}\label{10}
\begin{aligned}
S_{l_{jk}m_{jk}}(\mathbf{\hat{r}}) ={} & \sqrt{\Big( \frac{m_{jk}}{2|m_{jk}|} \Big)} \Bigg\{  Y_{l_{jk}-|m_{jk}|}(\mathbf{\hat{r}})+ \\
&  (-1)^m_{jk} \Big( \frac{m_{jk}}{|m_{jk}|} \Big)Y_{l_{jk}|m_{jk}|}(\mathbf{\hat{r}})  \Bigg\},
\end{aligned}
\end{equation}

and $m_{jk} = 0$: \hskip2ex
$S_{l_{jk}0}(\mathbf{\hat{r}}) = Y_{l_{jk}0}(\mathbf{\hat{r}}). \hskip 12ex \inlineeqno$ \label{11}

Here $Y_{lm}(\hat{\mathbf{r}})$ is the complex spherical harmonics.

Using the ortho-normalization property of the rotation matrix and integrating over the Euler angles, the TDCS is given as;
\begin{equation}\label{12}
\frac{d^3\sigma}{d\Omega_e d\Omega_s dE_e} = \frac{k_e k_s}{k_i} \sum_{k = 1} ^{N_j} \frac{a_{jk}^2}{\hat{l}_{jk}}\sum_{\mu = -l_{jk}}^{l_{jk}} |T_{fi}^{pw}(\mathbf{q})|^2,
\end{equation}
where $\hat{l}_{jk}$ = $2l_{jk}+1$.

We extend the formalism mentioned above for the twisted electron beam ionization for a molecular target in the coplanar asymmetric geometry by considering that the incident twisted electron beam propagates along the {\it z}-axis. Unlike plane wave, the incident momentum vector $\mathbf{k}_i$ of the twisted electron beam also has momentum distribution in the transverse direction and an inhomogeneous intensity distribution \cite{Dhankhar2022_2}. The following expression gives the incident momentum vector,
\begin{equation}\label{13}
\mathbf{k}_i = (k_i\sin{\theta_p}\cos{\phi_p})\hat{x}+(k_i\sin{\theta_p}\sin{\phi_p})\hat{y}+(k_i\cos{\theta_p})\hat{z}.  
\end{equation}
Here, \thetap \ and $\phi_p$ are the polar and azimuthal angles of the momentum vector, respectively. $\mathbf{k}_i$ carves out the surface of a cone when we change $\phi_p$ (for details see ref. \cite{Dhankhar2022}). The longitudinal component of the momentum $\mathbf{k}_{iz}$ is fixed, but the transverse momentum component $\mathbf{k}_{i\bot}$ has a direction that depends on $\phi_p$. It implies that the direction of incident momentum is not well defined. However, the magnitude of the transverse momentum is fixed and is given by $|\mathbf{k_{i\bot}}| = \varkappa = \sqrt{(k_i)^2 - (k_{iz})^2}$. By defining the {\it z}-axis along the longitudinal incident momentum transfer direction, the impact parameter $\mathbf{b}$, measures the transverse orientation of the target with respect to the axis of the incident beam.\\
The TDCS for a twisted electron is determined by calculating the scattering amplitude $T_{fi}^{tw}(\mathbf{q})$ and assuming that the outgoing electrons are detected with respect to the target. In momentum space representation, the Bessel beam (incident twisted electron beam) represents a superposition of plane waves, making an angle \thetap \ (also known as the opening angle) with the {\it z}-axis, a $\phi$ dependent phase ($e^{im\phi}$). The incident twisted electron wave function is thus expressed as,
\begin{equation}\label{14}
\Psi^{tw}_{\varkappa m}(\mathbf{r_i}) = \int^{2\pi}_{0}\frac{d\phi_p}{(2\pi)^2}a_{\varkappa m}(k_{i\bot})e^{i\mathbf{k}_i\cdot \mathbf{r_i}}, 
\end{equation}
where $\mathbf{r}_i$ is the position vector of the incident electron beam. When we replace the incident plane wave with the twisted wave in the theoretical formalism for the computation of TDCS for an (e, 2e) process of the molecular target (assuming that the target is located along the direction of incident twisted electron beam, $\mathbf{b}$ = 0), we get the following expression for twisted wave matrix element
\begin{equation} \label{15}
T^{tw}_{fi}(\varkappa,\mathbf{q,b}) = (-i)^m\int^{2\pi}_0\frac{d\phi_p}{2\pi}e^{im\phi_p}T^{pw}_{fi}(\mathbf{q}). 
\end{equation}
As evident from eqn. (\ref{15}), the twisted wave matrix element $T^{tw}_{fi}$ is expressed in terms of the plane wave matrix element $T^{pw}_{fi}$ \cite{Dhankhar2020}. The key difference here is that the momentum transfer vector $\mathbf{q} = \mathbf{k}_i-\mathbf{k}_s$ has to be calculated using the twisted wave momentum vector $\mathbf{k}_i$ (see eqn. (\ref{13})).
\begin{equation} \label{16}
q^2 = k^2_i + k^2_s -2k_ik_s\cos(\theta),  
\end{equation}
where,
\begin{equation} \label{17}
\cos(\theta) = \cos(\theta_p)\cos(\theta_s) + \sin(\theta_p)\sin(\theta_s)\cos(\phi_p-\phi_s).
\end{equation}
Here, $\phi_s$ is the azimuth angle of the scattered electron momentum vector $\mathbf{k}_s$. Unlike the plane wave, the momentum transfer of a twisted wave is not constant for a particular direction of $\mathbf{k}_s$ and depends on the azimuthal angle of the incident wave vector $\mathbf{k}_i$. This inherent uncertainty of momentum transfer direction for a twisted wave is accounted for by taking an integral over the azimuthal angle $\phi_p$ in eqn. (\ref{15}).
The TDCS for the molecular orbitals of \CH \ and \NH \ targets by twisted electron can be computed from eqn. (\ref{1}) together with the transition amplitude $T_{fi}^{tw}( \mathbf{q})$ from the eqn. (\ref{15}).

\subsection{Average over the impact parameter}
The process of ionization of a single molecule by a vortex beam is challenging experimentally. Therefore, the ionization process on a macroscopic target is preferable in a more realistic scenario. The cross-section for such a target can then be computed by taking the average of the plane wave cross-sections over all the possible impact parameters, \textbf{b}, in the transverse plane of the twisted electron beam.
The average cross-section, (TDCS)$_{av} = \overline{\frac{d^3 \sigma}{d\Omega_s d\Omega_e dE_e}}$ in terms of plane wave cross-section can be described as (for detailed derivation see  \cite{Serbo2015, Karlovets2017, Harris2019});
\begin{equation}\label{18}
(TDCS)_{av} = \frac{1}{2 \pi \cos\theta_p} \int_{0}^{2\pi} d\phi_p \frac{d^3 \sigma(\mathbf{q})}{d\Omega_s d\Omega_e dE_e},
\end{equation}
where $\frac{d^3 \sigma(\mathbf{q})}{d\Omega_s d\Omega_e dE_e}$ is like the TDCS for the plane wave electron beam depending on \textbf{q}.
From eqn. (\ref{18}), it is evident that the cross-section for the scattering of the twisted electrons by the macroscopic target is independent of the OAM number {\it m} of the incident twisted electron beam. However, (TDCS)$_{av}$ depends on the incident twisted electron beam's opening angle $\theta_p$.

\subsection{Superposition of {\it two} Bessel beams}
From eqn. \ref{18}, we see that the TDCS for a macroscopic molecular target is independent of the projection of the OAM number \m \ and the phase structure ($e^{im\phi}$) of the incident twisted electron beam. However, for a macroscopic target, we can restore the OAM sensitivity of the TDCS by considering the incident twisted electron beam as a superposition of the two beams with the same kinematic parameters, but different \m \ \cite{Serbo2015, Karlovets2017}. The following wave function describes a superposed twisted electron beam,
\begin{equation}\label{19}
\Psi(\mathbf{r}) = c_1 \Psi_{m_1}(\mathbf{r}) + c_2 \Psi_{m_2}(\mathbf{r}),
\end{equation}
where $\Psi_{m}(\mathbf{r})$ is given by eqn. \ref{14}, $c_n$ are the expansion coefficients given as
\begin{equation}\label{20}
c_n = |c_n|e^{i\alpha_n}, \hskip 2ex |c_1|^2 +|c_2|^2 = 1.
\end{equation}
Using eqn. \ref{19} in eqn. \ref{18}, we get the following expression for the TDCS \cite{Serbo2015, Karlovets2017}
\begin{equation}\label{21}
(TDCS)_{av} = \frac{1}{2 \pi \cos\theta_p} \int_{0}^{2\pi} d\phi_p G(\phi_p, \Delta m,\Delta \alpha) \frac{d^3 \sigma(\mathbf{q})}{d\Omega_s d\Omega_e dE_e} ,
\end{equation}
where $\Delta m$ = $m_2 - m_1$ is the difference in the OAM projections, $\Delta \alpha$ = $\alpha_2 - \alpha_1$ is the difference in the phases of the twisted states, and the factor 
$G(\phi_p, \Delta m,\Delta \alpha) = 1 + 2|c_1c_2| cos[(m_2-m_1)(\phi - \pi/2) + \alpha_2 - \alpha_1]$.
\section{Results and discussions}\label{sec3}
In this section, we present the results of our calculations for the single ionization of \CH \ and \NH \ by a twisted electron beam impact. As mentioned in Sec \ref{sec2}, we have used the theoretical model given in \cite{Dhankhar2022} for the single ionization of \CH  \ and \NH \ molecules by twisted electron impact. We compare the twisted beam TDCS with the plane wave TDCS by keeping the opening angle \thetap \ the same as the scattering angle \thetas \ for different values of OAM number \m. We have used the kinematics; scattered energy ($E_s$) = 500eV, ejected energy ($E_e$) = 74eV and scattering angle (\thetas) = 6\td \ in the coplanar geometry \cite{LB2009,Nixon2013}. As mentioned in Sec \ref{sec1} and \ref{sec2}, the molecular orbitals of \CH \ and \NH \ have \p-type and \s-type characters. To study the dynamics of the twisted electron impact ionization of these molecules, we discuss the results of TDCSs for the orbitals of \p-type and \s-type characters separately.  
\subsection{Ionization from orbitals of {\it p}-type character of the targets}\label{sec3.1}
\begin{figure}[h]
\centering
\includegraphics[width=1.0\columnwidth]{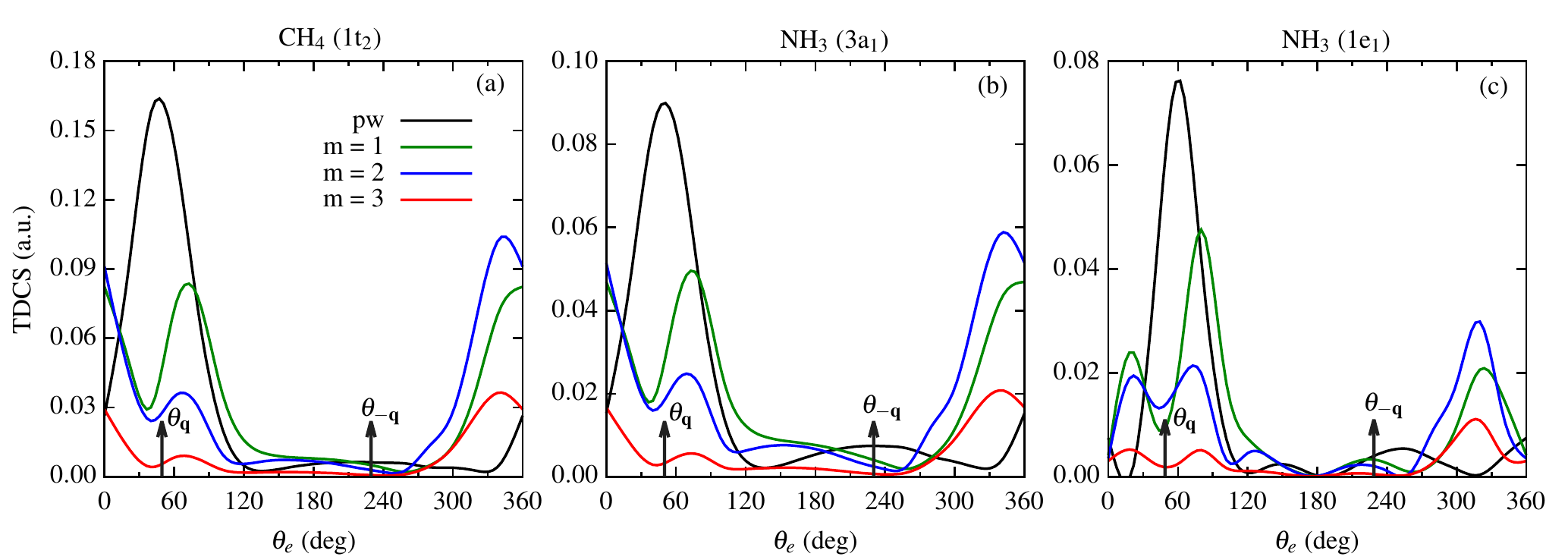}
\caption{Angular profiles of the TDCS as a function of the ejected electron angle \thetae \  for the twisted electron (e, 2e) process from the orbitals of \p-type character of \CH \ and \NH \ in the coplanar asymmetric geometry. The kinematics used here is $E_s$ = 500eV, $E_e$ = 74eV, \thetas \ = 6\td \ and \thetap \ = \thetas. The sub-figure (a) is for the 1t$_2$ orbital of \CH \ molecular target, while the sub-figures (b) and (c) are for the 3a$_1$ and 1e$_1$ molecular orbitals of the \NH \ molecule, respectively. The black, green, blue, and red curves represent the TDCS for plane wave, \m \ = 1, 2, and 3, respectively. For all the sub-figures, we have used the multiplicative factors 10, 40, and 50 for \m \ = 1, 2, and 3, respectively. \thetaq \ and \tq \ represent the direction of momentum transfer and recoil momentum.}\label{fig1}
\end{figure}

Figure \ref{fig1} represents the TDCS for the twisted electron impact ionization of the molecular targets \CH \ and \NH \ for the orbitals having {\it p}-type character. We present the TDCS as a function of the ejected electron angle in the coplanar asymmetric geometry for \CH \ (1t$_2$) and \NH \ (3a$_1$ and 1e$_1$). For a central collision $|\mathbf{b}|$ = 0, we study the effect of OAM number \m \ on the angular profiles of the TDCS for \thetap \ = \thetas. In figure \ref{fig1}, black, green, blue, and red curves represent the TDCS for plane wave, OAM number \m \ = 1, 2, and 3 respectively for different molecular orbitals (having \p-type character) as mentioned in the frames.

We use the scaling factors 10, 40, and 50 for the TDCS in figure \ref{fig1} for \m \ =  1, 2, and 3, respectively, to compare them with the plane wave TDCS. For the molecular orbitals considered in figure \ref{fig1}, we observe a binary and recoil peak structure in the angular profiles of the TDCS along the momentum transfer directions \thetaq \ and \tq \ for the plane wave. For the twisted electron beam (e, 2e) process, the angular profiles differ from the plane wave TDCS profiles. We observe that the magnitude of the TDCS decreases with an increase in the OAM number \m \ for all the cases (for explanation describing the decrease in magnitude for fixed \thetap \ and variable \m \ see \cite{Dhankhar2022}). For the outermost molecular orbitals of \CH \ and \NH, we observe peaks in the binary and forward regions in the angular profiles with a minimum around the momentum transfer direction (see blue, green, and red curves in figure \ref{fig1} (a) and (b)). The peaks, however, in the binary region are shifted from the momentum transfer direction (see blue, green, and red curves in figure \ref{fig1} (a) and (b) around \thetaq). 

The angular profiles of the twisted electron TDCS for the \p-type orbitals of \CH \ and \NH \ molecules depict a two-peak structure (or peak splitting) in the binary region. This two-peak structure is a characteristic of the \p-type orbitals (see green, blue, and red curves in figure \ref{fig1} around the momentum transfer direction \thetaq). As mentioned in \cite{LB2009, Dhankhar2022_2}, the splitting of the binary peak is a characteristic of the \p-type orbitals and depends on the kinematical conditions. This splitting of the binary peak in the TDCS for the plane wave (e, 2e) process is explained by the Bethe-Ridge condition. According to Bethe-Ridge \cite{Khajuria1999}, when the recoil momentum $\mathbf{q}_r$ = $\mathbf{q}$ - $\mathbf{k}_e$ is minimum, the binary peak splits. The position of \thetae \ at which $\mathbf{q}_r$ is minimum results in a vanishing cross-section at that angle, leading to the peak splitting at that \thetae. With the present kinematics for the plane wave (e, 2e) on these \p-type orbitals, we only obtain a binary and recoil peak structure in the angular profile of the TDCS. However, we observe splitting in the binary peak for the twisted electron impact ionization at the present kinematics. This splitting of the binary peak for the twisted (e, 2e) process is similar to the one observed for Ne (2p) and Ar (3p) in \cite{Dhankhar2022_2}. However, in \cite{Dhankhar2022_2}, the splitting was observed for opening angles higher than \thetas \, and here it is at \thetap \ = \thetas. In the present study, we observe splitting around momentum transfer direction for 1t$_2$ and 3a$_1$ (the outermost orbitals of \CH \ and \NH). However, the splitting is distributed around the forward direction, ({\it i.e.}, the peaks are located around \thetae \ = 0\td \ and 360\td \ in figure \ref{fig1}(a) and (b)). We observe a prominent peak splitting in the binary region for 1e$_1$ orbital (see figure \ref{fig1}(c)).

\subsection{Ionization from orbitals of \s-type character of the targets}\label{sec3.2}
\begin{figure}[h]
\centering
\includegraphics[width=1.0\columnwidth]{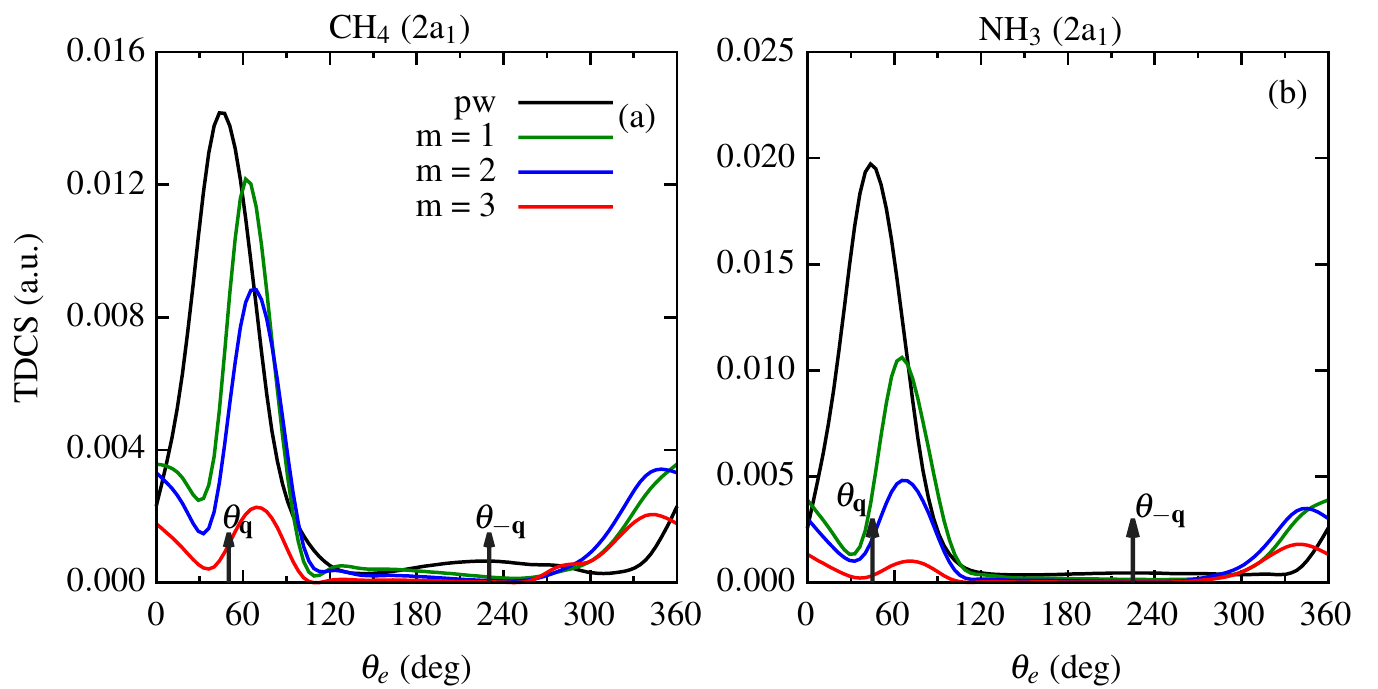}
\caption{Same as figure \ref{fig1}, except the angular profiles here are for the orbitals of \s-type character of \CH \ and \NH \ in the co-planar asymmetric geometry.}
\label{fig2}
\end{figure}

In this sub-section, we discuss the results of our calculations for the twisted electron impact ionization TDCS of the molecular targets \CH \ and \NH \ with the orbitals of \s-type character. We present the TDCS as a function of the ejected electron angle in the coplanar asymmetric geometry for \CH \ (2a$_1$) and \NH \ (2a$_1$). Here also, we keep \thetap \ = \thetas \ and study the angular profiles of the TDCS for different \m. 

The magnitude of the TDCS is smaller than the plane wave TDCS. We use the scaling factors 5, 15, and 30 for the TDCS in figure \ref{fig2} for \m \ =  1, 2, and 3, respectively, to compare their respective TDCSs with the plane wave TDCS. Here also, the angular profiles of the TDCS for plane wave depict a binary and recoil peak structure along the momentum transfer directions \thetaq \ and \tq. For the twisted electron impact ionization, we observe binary peak as the dominant peak with the peak position shifted from the momentum transfer direction for all \m  \ (see green, blue, and red curves around \thetae \ = \thetaq \ in figure \ref{fig2}). As observed for the \p-type orbitals (see figure \ref{fig1}), we also observe peaks in the forward direction (see green, blue, and red curves around \thetae \ = 0\td \ or 360\td \ region in figure \ref{fig2}). However, in this case, the peaks are not as prominent as they were for the outer orbitals (see green, blue, and red curves in figure \ref{fig1} and \ref{fig2} around \thetae \ = 360\td \ for comparison).

The above discussion makes it clear that in a twisted electron impact ionization process, there is a transfer of the intrinsic OAM of the beam, and angular profiles of the TDCS show a significant dependence on the OAM number \m \ for \thetap \ = \thetas.

\subsection{Angular profiles for the (TDCS)$_{av}$ for the macroscopic molecular targets}\label{sec3.3}
In figures \ref{fig3} and \ref{fig4}, we present the angular profiles of the TDCS averaged over the impact parameter $\mathbf{b}$, \TDCSAV, for the \CH \ and \NH \ molecular targets. We discuss the results of the \TDCSAV \ as a function of the ejected electron angle \thetae \ for \p-type and \s-type orbitals separately. The \TDCSAV \ depends on the opening angle \thetap \ of the incident twisted electron beam. Figures \ref{fig3} and \ref{fig4} represent the \TDCSAV \ for the kinematics $E_s$ = 500eV, $E_e$ = 74eV and \thetas \ = 6\td. We present the results of the \TDCSAV \ for the opening angles 1\td, 6\td, 15\td, and 20\td \ for both molecular targets. 

\subsubsection{(TDCS)$_{av}$ from orbitals of {\it p}-type character of the targets}\label{avb_p}
\begin{figure}[h]
\centering
\includegraphics[width=1.0\columnwidth]{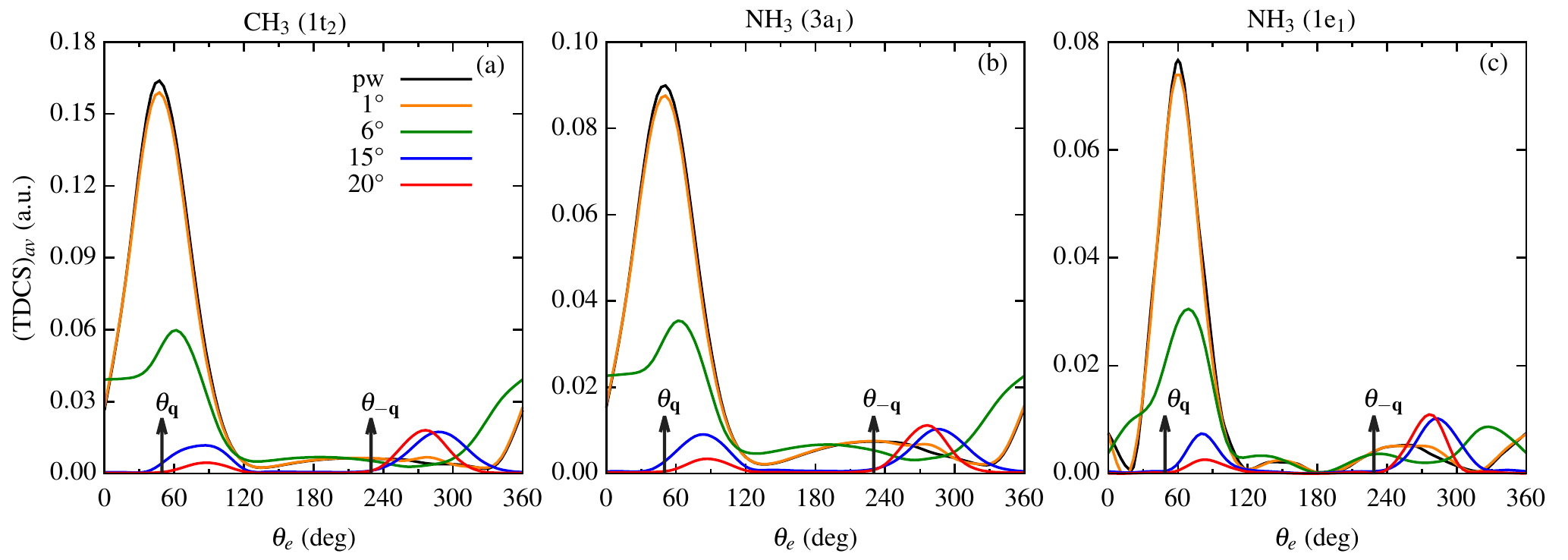}
\caption{(TDCS)$_{av}$ plotted as a function of the ejected electron angle \thetae \ for the plane wave (solid line) and twisted electron beam for different opening angles as shown in the frames of each sub-figure. The kinematics is the same as in Fig \ref{fig1}. The black, orange, green, blue, and red curves represent the TDCS for plane wave, \thetap \ = 1\td, 6\td, 15\td, and 20\td, respectively.}\label{fig3}
\end{figure}

In this sub-section, we present the results of the \TDCSAV \ from the orbitals of \p-type character of \CH \ and \NH \ molecular targets, {\it i.e.} 1t$_2$ of \CH \ and 3a$_1$ and 1e$_1$ of \NH. 

From fig \ref{fig3}, we observe that for the smallest opening angle, such as \thetap \ = 1\td, the angular profile of the \TDCSAV \ is similar to that of the plane wave for all the cases (see orange curves in figure \ref{fig3}). For \thetap \ = \thetas \ = 6\td, the peak in the binary region shifts from the momentum transfer direction for all the cases. However, for the outermost orbitals (1t$_2$ of \CH \ and 3a$_1$ of \NH), we observe peaks in the binary and backward region (see green curves in figure \ref{fig3} (a) and (b) around \thetae = \thetaq \ and \thetae \ = 180\td). For the 1e$_1$ orbital of \NH \ molecule, we observe that the binary peak shifts from the momentum transfer direction and the small peak around the recoil momentum direction (see the green curve in figure \ref{fig3} (c) around the arrows). On further increasing the opening angles to 15\td \ and 20\td, the angular profiles depict a two-peak structure with peaks in the perpendicular directions (see blue and red curves in figure \ref{fig3} around \thetae \ = 90\td and \ 270\td). The magnitude of the \TDCSAV \ also decreases with an increase in the opening angle. 

\subsubsection{(TDCS)$_{av}$ from orbitals of \s-type character of the targets}\label{avb_s}
\begin{figure}[h]
\centering
\includegraphics[width=1.0\columnwidth]{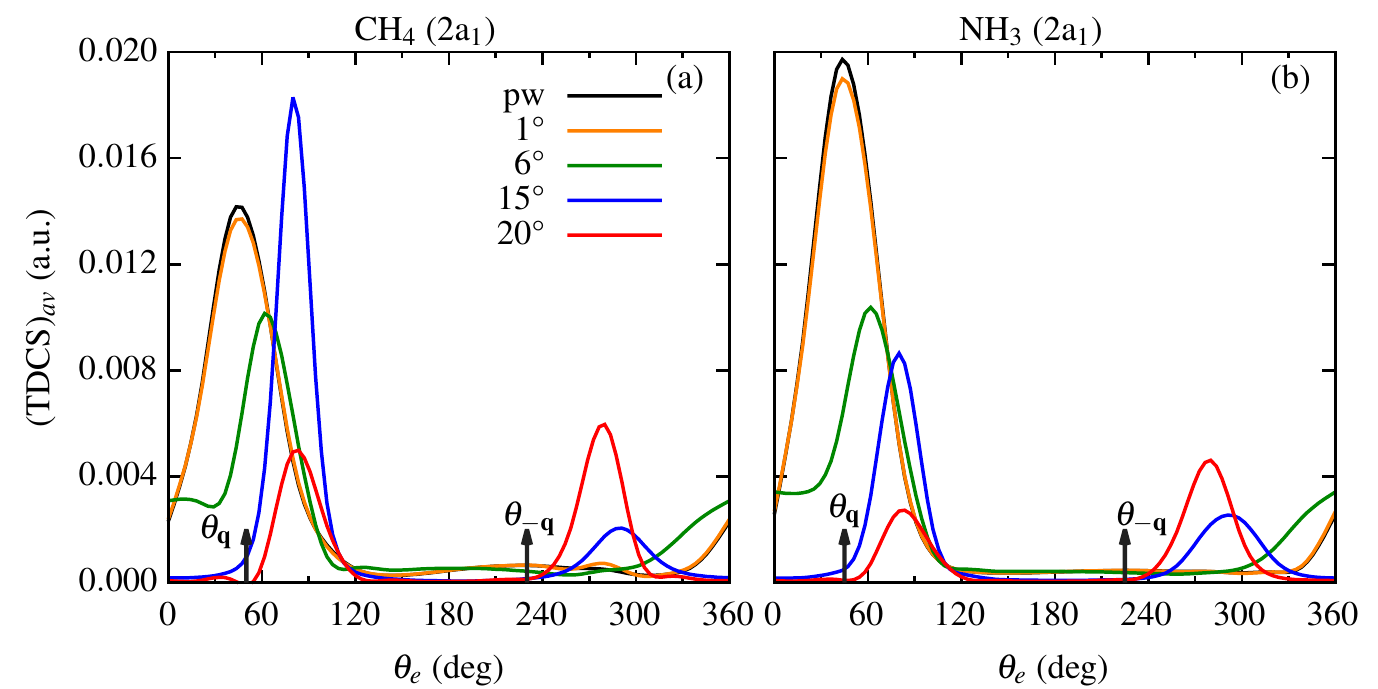}
\caption{(TDCS)$_{av}$ plotted as a function of the ejected electron angle \thetae \ for the plane wave (solid line) and twisted electron beam for different opening angles and molecular orbitals as shown in the frames of each sub-figure. The kinematics is same as used in Fig \ref{fig2}.}\label{fig4}
\end{figure}
In this sub-section, we present the results of the \TDCSAV \ from the orbitals of \s-type character of \CH \ and \NH \ molecular targets, {\it i.e.} 2a$_1$ of \CH \ and 2a$_1$ of \NH. 
Similar to figure \ref{fig3}, for the \s-type orbitals of \CH \ and \NH \ also, the angular profiles of the \TDCSAV \ for the smallest opening angle are similar to the plane wave profile (see orange curves in figure \ref{fig4}). The angular profiles for the opening angle, same as the scattering angle, ({\it i.e.} \thetap \ = \thetas \ = 6\td), depict a dominant one peak structure in the binary region with the peak shifted from the momentum transfer direction (see green curves in figure \ref{fig4}). For the \s-type orbitals, some interesting features are observed at higher opening angles. For the 2a$_1$ orbital of \CH \ molecule, for opening angle 15\td, the magnitude of the \TDCSAV \ is higher than that of the plane wave. In our previous study, we observed a similar phenomenon on the twisted electron impact ionization of water molecule \cite{Dhankhar2022}. However, in \cite{Dhankhar2022}, we observed that the peaks have a higher magnitude for \thetap \ = \thetas \ case and for \p-type orbitals only. For the 2a$_1$ orbital of \NH \ molecule, the magnitude of the \TDCSAV \ is less than the plane wave TDCS. However, unlike the \p-type orbitals, here for \thetap \ = 15\td \ prominent peak is in the perpendicular direction (compare blue curves in figure \ref{fig3} and \ref{fig4}). On further increasing the opening angle, the magnitude of the \TDCSAV \ decreases, and the peaks shift toward the perpendicular direction (see red curves in figure \ref{fig4}).

From the sub-sections \ref{avb_p} and \ref{avb_s}, we conclude that the angular profiles of the \TDCSAV \ depend on the opening angle \thetap.

\subsection{TDCS from the coherent superposition of Bessel beams}\label{sec3.4}

\begin{figure}[h]
\centering
\includegraphics[width=1.0\columnwidth]{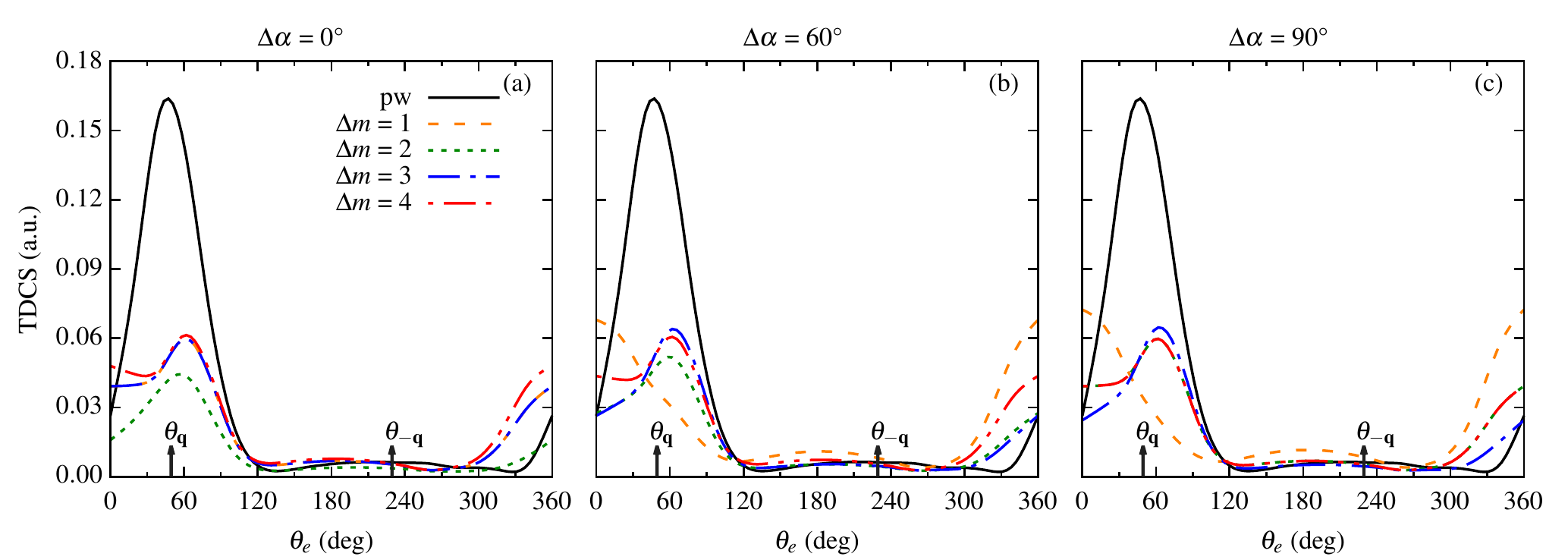}
\caption{TDCS as a function of ejected electron angle \thetae \ for the twisted electron (e, 2e) process on the 1t$_2$ orbital of \CH \ molecular target. The results are for a coherent superposition of two Bessel beams with different OAM projections and phases (as shown in the sub-figures). We keep the kinematics the same as figure \ref{fig1}. The solid, dashed, dotted, dashed-dotted and dashed-dotted-dotted curves are for plane wave, \Delm \ = 1, 2, 3 and 4 respectively.}\label{fig5}
\end{figure}

\begin{figure}[h]
\centering
\includegraphics[width=1.0\columnwidth]{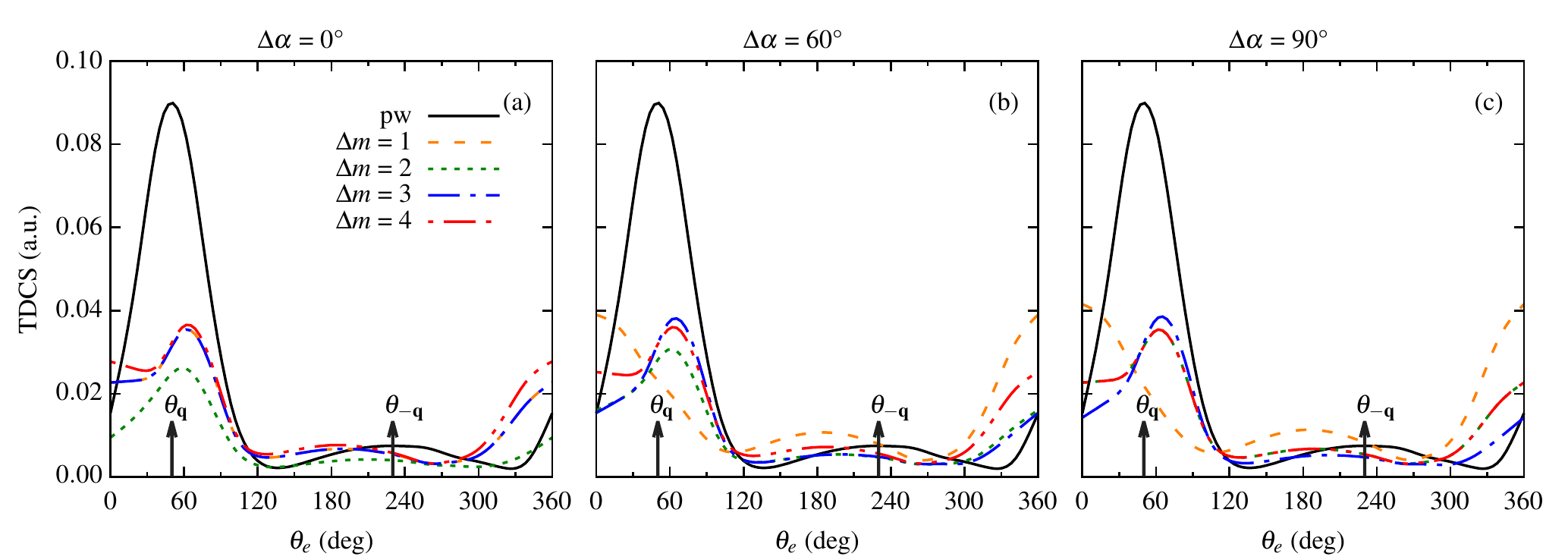}
\caption{Same as figure \ref{fig5}, except the results are for 3a$_1$ molecular orbital of \NH.}\label{fig6}
\end{figure}
\begin{figure}[h]
\centering
\includegraphics[width=1.0\columnwidth]{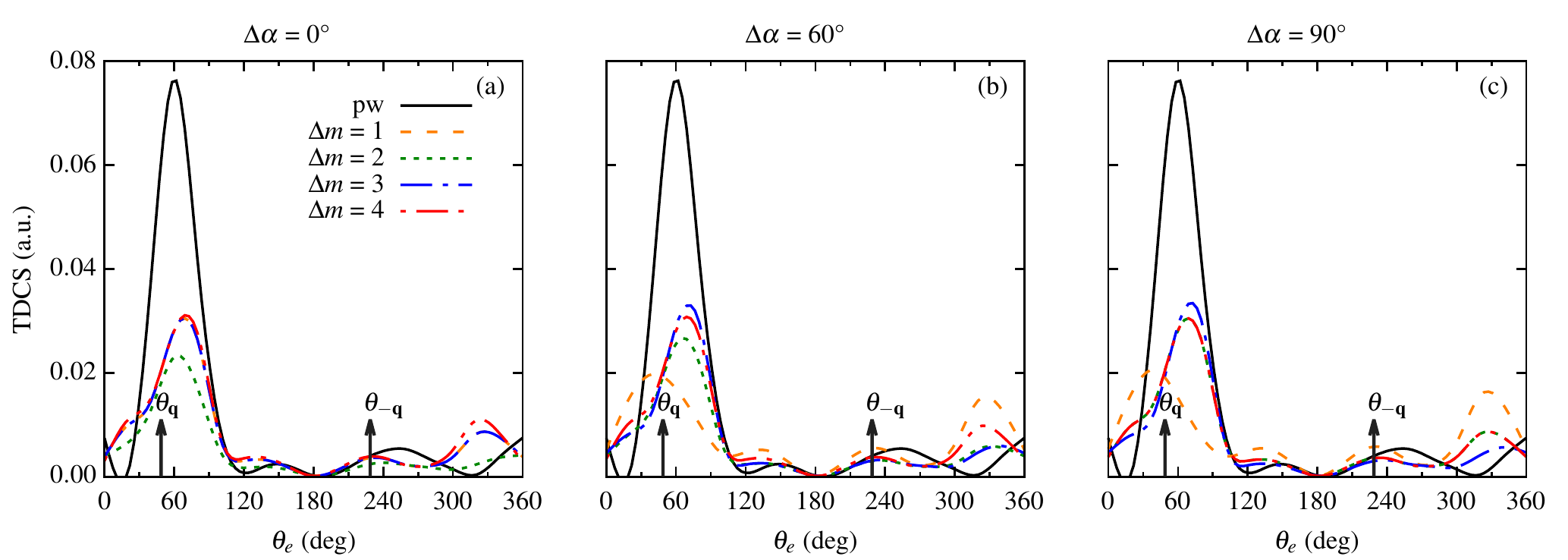}
\caption{Same as figure \ref{fig5}, except the results are for 1e$_1$ molecular orbital of \NH.}\label{fig7}
\end{figure}
\begin{figure}[h]
\centering
\includegraphics[width=1.0\columnwidth]{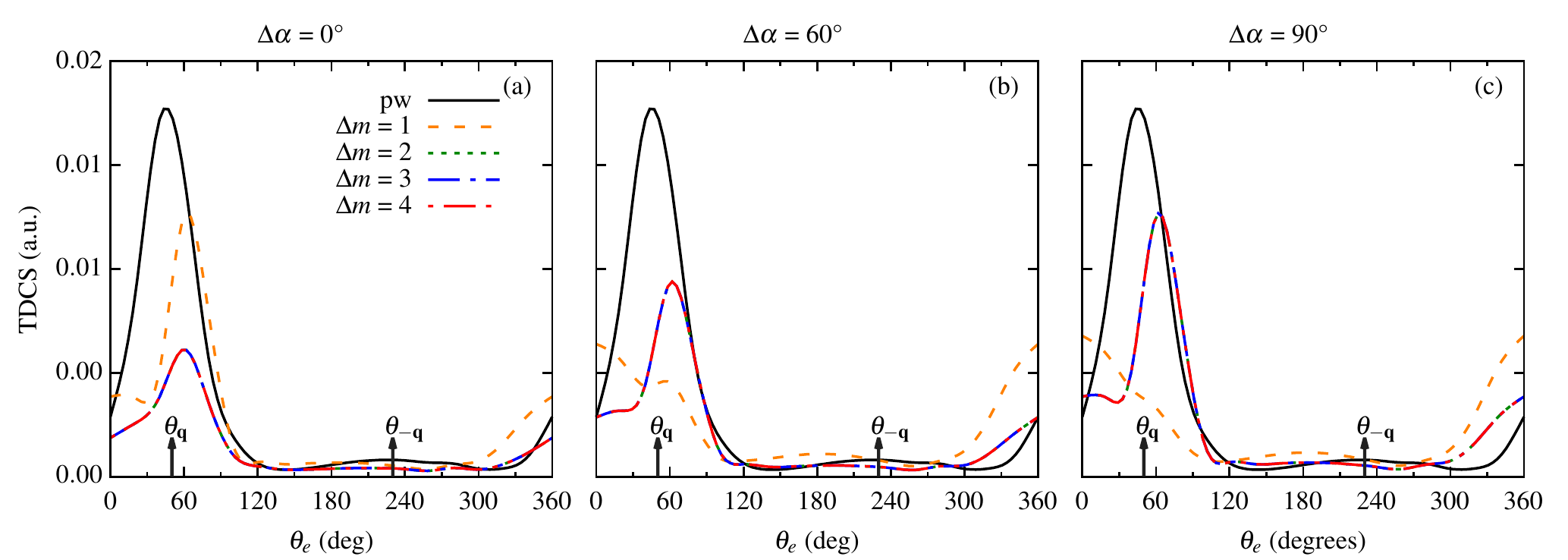}
\caption{Same as figure \ref{fig5}, except the results are for 2a$_1$ molecular orbital of \CH.}\label{fig8}
\end{figure}
\begin{figure}[h]
\centering
\includegraphics[width=1.0\columnwidth]{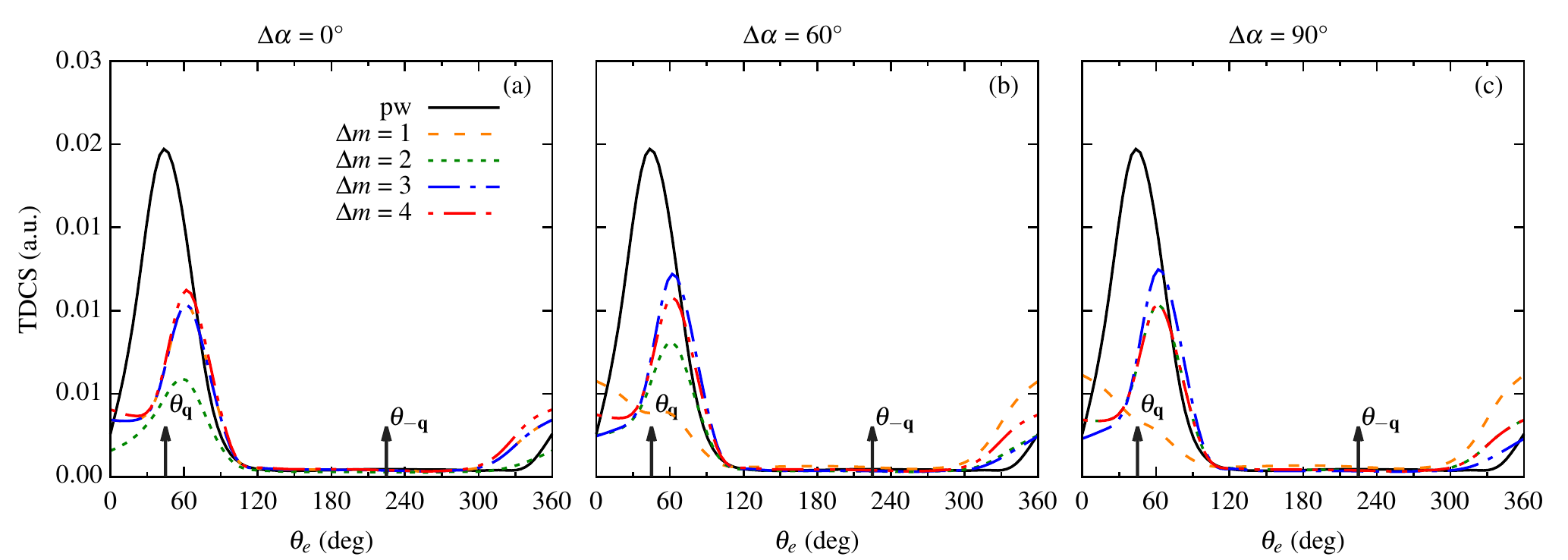}
\caption{Same as figure \ref{fig5}, except the results are for 2a$_1$ molecular orbital of \NH.}\label{fig9}
\end{figure}

In the previous section, we discussed the twisted electron ionization from the macroscopic molecular target. For such a target, the averaged TDCS is independent of the OAM projection \m. However, one can restore and study the OAM sensitivity by considering the incident electron beam as a coherent superposition of {\it two} twisted states with different \m's. The advantage of such a superimposed Bessel beam is that it helps to investigate the effect of OAM projection \m \ on the (e, 2e) process using a twisted electron beam for a realistic scenario. The TDCS now depends not only on the opening angle \thetap but also on the OAM projections \Delm \ and the difference in the beam phases \Dela. In figures \ref{fig5}-\ref{fig9}, we present the results of our calculations for the twisted electron impact ionization from {\it two} coherent Bessel beams for the \CH \ and \NH \ molecular targets. We present the results for different \Delm \ (1, 2, 3 and 4)  and \Dela \ (0\td, 60\td \ and 90\td)  with the same kinematics as used in figures \ref{fig1} and \ref{fig2}. Like the previous sections, we discuss the angular profiles \p-type and \s-type orbitals. 

Figures \ref{fig5}-\ref{fig7} represent the TDCS from the coherent superposition of {\it two} Bessel beams for the molecular orbitals of \p-type character, namely,  1t$_2$ (\CH), 3a$_1$ (\NH) and 1e$_1$ (\NH) respectively. For \Dela \ = 0\td, {\it i.e.} when both the incident twisted electron beams are in phase, we observe peaks in the binary and recoil regions with peak positions shifted from the momentum transfer direction (see figure \ref{fig5}-\ref{fig7} (a)) for the three \p-dominant orbitals. Also, the magnitude of TDCS for all the \Delm s is less than the plane wave magnitude, with \Delm \ = 2 having the smallest magnitude. As we increase the phase between the Bessel beams, from 0\td \ to 60\td, the angular profiles for \Delm \ = 2, 3, and 4 remain almost identical to that of \Dela \ = 0\td \ with a small increment in magnitude (see dotted, dashed-dotted and dashed-dotted-dotted curves in figures \ref{fig5}-\ref{fig7}). However, for the outer orbitals of \CH \ (1t$_2$) and \NH \ (3a$_1$) and \Delm \ = 1, the angular profiles depict peaks in the forward and backward regions (see dashed curves in figures \ref{fig5} and \ref{fig6} around \thetae \ = 0\td(360\td) and 180\td). For \Delm \ = 1, we observe peaks in the binary and recoil regions with the lowest magnitude of TDCS. On further increasing the phase between incident beams to 90\td, we observe that the angular profile structure remains the same as that for \Delm \ = 60\td, but there is a slight increase in the magnitude for all the OAM projections (see dashed curves in sub-figures \ref{fig5}-\ref{fig7} (b) and (c)). 

Figures \ref{fig8}-\ref{fig9} represent the TDCS from the coherent superposition of {\it two} Bessel beams for the molecular orbitals of \s-type character, namely,  2a$_1$ (\CH) and 2a$_1$ (\NH) respectively. For the 2a$_1$ orbital of \CH \ molecule, we observe the binary and recoil peak structure with the same magnitude of TDCS for \Delm \ = 2, 3, and 4 at different phases of the incident twisted electron beams as shown in the figure \ref{fig8}. For these values of \Delm, the peak position shifts to a perpendicular direction with an increase in the phase between the incident twisted electron beams (see dotted, dashed-dotted, and dashed-dotted-dotted curves in figure \ref{fig8} (b) and (c)). For the in-phase incidence of the twisted electron beams, the magnitude of the TDCS for \Delm \ = 1 is highest compared to the other \Delm s. The binary peak structure disappears with a further increment in the phase between the incident beams. We observe peaks in the forward and directions (see dashed curves for different \Dela s in figure \ref{fig8}).

The results of our calculations for the 2a$_1$ orbital of the \NH \ molecule for superposed Bessel beams are pretty similar to the ones observed for the \p-type orbitals (compare figure \ref{fig5} and \ref{fig9}). For \Dela \ = 0\td, the angular profile of TDCS maintains the binary peak structure of the plane wave for the \Delm s considered, with \Delm \ = 2 having the lowest magnitude. On further increasing the phase between incident beams, \Delm = 2, 3, and 4  maintains the binary peak structure of the angular profile with peak shifted from the momentum transfer direction and an enhancement in the magnitude of TDCS (see dotted, dashed-dotted and dashed-dotted-dotted curves in figure \ref{fig9}). However, for \Delm \ = 1, the binary peak structure in the angular profile of the TDCS disappears with an increment in the phase angle, and we observe peaks in forward and backward directions (see dashed curves in figure \ref{fig9} around \thetae \ = 0\td(360\td) and 180\td).

\section{Conclusions}\label{sec4}
We have presented the theoretical study of the triple differential cross-sections (TDCS) for an (e, 2e) process for \CH \ and \NH \ molecular targets by the twisted electron beam. We studied the angular distributions of the TDCS for the coplanar asymmetric geometry in the first Born approximation for the twisted electron beam. We have studied the influence of the OAM number \m \ on the angular profiles of the TDCS for the opening angle, the same as the scattering angle (\thetap \ = \thetas). In this paper, we have extended our previous model for the twisted electron impact ionization for the water molecule \cite{Dhankhar2022} to the \CH \ and \NH \ molecular targets. We have studied the angular profiles for the \p-type and \s-type molecular orbitals of \CH \ and \NH. For the present geometry, the angular profiles of the TDCS (for both \p-type and \s-type) peaks around $\theta_\mathbf{q}$ (binary peak) and $\theta_{-\mathbf{q}}$ (recoil peak) for the plane wave. However, the angular profiles of the TDCS show significant deviation from the plane wave TDCS angular profiles for the molecular orbitals considered. The twisted electron TDCS retrieves the characteristic two-peak structure for the 1e$_1$ orbital of \NH \, which is absent in the plane wave angular distributions of the TDCS. For the \s-type orbitals, the angular distributions are shifted from the momentum transfer direction.

For a macroscopic molecular target, we discuss the (TDCS)$_{av}$ (averaged over the impact parameter \textbf{b}) as a function of the opening angle $\theta_p$ of the twisted electron beam. The results show that the angular profiles of (TDCS)$_{av}$ significantly depend on the opening angle ($\theta_p$) of the twisted electron beam.
The (TDCS)$_{av}$ is similar to the plane wave TDCS for a smaller opening angle (\thetap \ = 1\td). For opening angles higher than the scattering angle, {\it i.e.} 15\td \ and 20\td, the peaks in the angular profiles shift to perpendicular directions. We also discussed the angular profiles for the coherent superposition of {\it two} Bessel beams. The angular profiles for superposed Bessel beams show the dependence of the TDCS on the OAM projection \m. 

Our present investigation of \CH \ and \NH \ molecule is based on our previous studies on H$_2$O. In our earlier study \cite{Dhankhar2022}, we focused on the angular profiles of the TDCS for different molecular orbitals of the water molecule, which constituted mainly {\it p}-type orbitals. The angular profiles of the TDCS for an incident plane wave depicted peak splitting in the {\it p}-type orbitals, while there was no peak splitting for angular profiles of the twisted electron beam.
In the present communication, we study the cross-sections for both {\it s} and {\it p}-type orbitals here. The angular profiles for the twisted electron beam, as shown in \ref{sec3}, exhibit significantly different observations than those we studied earlier. Like,  the splitting of the prominent peaks for {\it p}-type orbitals for the twisted electron case contrary to the $H_2O$ TDCS (where we did not observe the splitting in the twisted electron case for \thetap \ = \thetas \ case), a significant deviation of the angular profiles of the (TDCS)$_{av}$ for the opening angle other than $\theta_p$ = $\theta_s$, etc. The results suggest that the twisted electron beams present a better tool than the plane-wave beam to image the electron cloud and study the molecular structure.



\begin{thebibliography}{51}%
	\makeatletter
	\providecommand \@ifxundefined [1]{%
		\@ifx{#1\undefined}
	}%
	\providecommand \@ifnum [1]{%
		\ifnum #1\expandafter \@firstoftwo
		\else \expandafter \@secondoftwo
		\fi
	}%
	\providecommand \@ifx [1]{%
		\ifx #1\expandafter \@firstoftwo
		\else \expandafter \@secondoftwo
		\fi
	}%
	\providecommand \natexlab [1]{#1}%
	\providecommand \enquote  [1]{``#1''}%
	\providecommand \bibnamefont  [1]{#1}%
	\providecommand \bibfnamefont [1]{#1}%
	\providecommand \citenamefont [1]{#1}%
	\providecommand \href@noop [0]{\@secondoftwo}%
	\providecommand \href [0]{\begingroup \@sanitize@url \@href}%
	\providecommand \@href[1]{\@@startlink{#1}\@@href}%
	\providecommand \@@href[1]{\endgroup#1\@@endlink}%
	\providecommand \@sanitize@url [0]{\catcode `\\12\catcode `\$12\catcode
		`\&12\catcode `\#12\catcode `\^12\catcode `\_12\catcode `\%12\relax}%
	\providecommand \@@startlink[1]{}%
	\providecommand \@@endlink[0]{}%
	\providecommand \url  [0]{\begingroup\@sanitize@url \@url }%
	\providecommand \@url [1]{\endgroup\@href {#1}{\urlprefix }}%
	\providecommand \urlprefix  [0]{URL }%
	\providecommand \Eprint [0]{\href }%
	\providecommand \doibase [0]{http://dx.doi.org/}%
	\providecommand \selectlanguage [0]{\@gobble}%
	\providecommand \bibinfo  [0]{\@secondoftwo}%
	\providecommand \bibfield  [0]{\@secondoftwo}%
	\providecommand \translation [1]{[#1]}%
	\providecommand \BibitemOpen [0]{}%
	\providecommand \bibitemStop [0]{}%
	\providecommand \bibitemNoStop [0]{.\EOS\space}%
	\providecommand \EOS [0]{\spacefactor3000\relax}%
	\providecommand \BibitemShut  [1]{\csname bibitem#1\endcsname}%
	\let\auto@bib@innerbib\@empty
	\bibitem [{\citenamefont {Torres}\ and\ \citenamefont
		{Torner}(2011)}]{Torres2011}%
	\BibitemOpen
	\bibfield  {author} {\bibinfo {author} {\bibfnamefont {J.~P.}\ \bibnamefont
			{Torres}}\ and\ \bibinfo {author} {\bibfnamefont {L.}~\bibnamefont
			{Torner}},\ }\href@noop {} {\emph {\bibinfo {title} {Twisted photons:
				applications of light with orbital angular momentum}}}\ (\bibinfo
	{publisher} {John Wiley \& Sons},\ \bibinfo {year} {2011})\BibitemShut
	{NoStop}%
	\bibitem [{\citenamefont {Molina-Terriza}, \citenamefont {Torres},\ and\
		\citenamefont {Torner}(2007)}]{Molina2007}%
	\BibitemOpen
	\bibfield  {author} {\bibinfo {author} {\bibfnamefont {G.}~\bibnamefont
			{Molina-Terriza}}, \bibinfo {author} {\bibfnamefont {J.~P.}\ \bibnamefont
			{Torres}}, \ and\ \bibinfo {author} {\bibfnamefont {L.}~\bibnamefont
			{Torner}},\ }\href {\doibase 10.1038/nphys607} {\bibfield  {journal}
		{\bibinfo  {journal} {Nat. Phys.}\ }\textbf {\bibinfo {volume} {3}},\
		\bibinfo {pages} {305} (\bibinfo {year} {2007})}\BibitemShut {NoStop}%
	\bibitem [{\citenamefont {Bliokh}\ \emph {et~al.}(2017)\citenamefont {Bliokh},
		\citenamefont {Ivanov}, \citenamefont {Guzzinati}, \citenamefont {Clark},
		\citenamefont {{Van Boxem}}, \citenamefont {B{\'{e}}ch{\'{e}}}, \citenamefont
		{Juchtmans}, \citenamefont {Alonso}, \citenamefont {Schattschneider},
		\citenamefont {Nori},\ and\ \citenamefont {Verbeeck}}]{Bliokh2017}%
	\BibitemOpen
	\bibfield  {author} {\bibinfo {author} {\bibfnamefont {K.}~\bibnamefont
			{Bliokh}}, \bibinfo {author} {\bibfnamefont {I.}~\bibnamefont {Ivanov}},
		\bibinfo {author} {\bibfnamefont {G.}~\bibnamefont {Guzzinati}}, \bibinfo
		{author} {\bibfnamefont {L.}~\bibnamefont {Clark}}, \bibinfo {author}
		{\bibfnamefont {R.}~\bibnamefont {{Van Boxem}}}, \bibinfo {author}
		{\bibfnamefont {A.}~\bibnamefont {B{\'{e}}ch{\'{e}}}}, \bibinfo {author}
		{\bibfnamefont {R.}~\bibnamefont {Juchtmans}}, \bibinfo {author}
		{\bibfnamefont {M.}~\bibnamefont {Alonso}}, \bibinfo {author} {\bibfnamefont
			{P.}~\bibnamefont {Schattschneider}}, \bibinfo {author} {\bibfnamefont
			{F.}~\bibnamefont {Nori}}, \ and\ \bibinfo {author} {\bibfnamefont
			{J.}~\bibnamefont {Verbeeck}},\ }\href {\doibase
		https://doi.org/10.1016/j.physrep.2017.05.006} {\bibfield  {journal}
		{\bibinfo  {journal} {Phys. Rep.}\ }\textbf {\bibinfo {volume} {690}},\
		\bibinfo {pages} {1} (\bibinfo {year} {2017})},\ \bibinfo {note} {theory and
		applications of free-electron vortex states}\BibitemShut {NoStop}%
	\bibitem [{\citenamefont {Lloyd}\ \emph {et~al.}(2017)\citenamefont {Lloyd},
		\citenamefont {Babiker}, \citenamefont {Thirunavukkarasu},\ and\
		\citenamefont {Yuan}}]{Lloyd2017}%
	\BibitemOpen
	\bibfield  {author} {\bibinfo {author} {\bibfnamefont {S.~M.}\ \bibnamefont
			{Lloyd}}, \bibinfo {author} {\bibfnamefont {M.}~\bibnamefont {Babiker}},
		\bibinfo {author} {\bibfnamefont {G.}~\bibnamefont {Thirunavukkarasu}}, \
		and\ \bibinfo {author} {\bibfnamefont {J.}~\bibnamefont {Yuan}},\ }\href
	{\doibase 10.1103/RevModPhys.89.035004} {\bibfield  {journal} {\bibinfo
			{journal} {Rev. Mod. Phys.}\ }\textbf {\bibinfo {volume} {89}},\ \bibinfo
		{pages} {035004} (\bibinfo {year} {2017})}\BibitemShut {NoStop}%
	\bibitem [{\citenamefont {Larocque}\ \emph {et~al.}(2018)\citenamefont
		{Larocque}, \citenamefont {Kaminer}, \citenamefont {Grillo}, \citenamefont
		{Leuchs}, \citenamefont {Padgett}, \citenamefont {Boyd}, \citenamefont
		{Segev},\ and\ \citenamefont {Karimi}}]{Hugo2018}%
	\BibitemOpen
	\bibfield  {author} {\bibinfo {author} {\bibfnamefont {H.}~\bibnamefont
			{Larocque}}, \bibinfo {author} {\bibfnamefont {I.}~\bibnamefont {Kaminer}},
		\bibinfo {author} {\bibfnamefont {V.}~\bibnamefont {Grillo}}, \bibinfo
		{author} {\bibfnamefont {G.}~\bibnamefont {Leuchs}}, \bibinfo {author}
		{\bibfnamefont {M.~J.}\ \bibnamefont {Padgett}}, \bibinfo {author}
		{\bibfnamefont {R.~W.}\ \bibnamefont {Boyd}}, \bibinfo {author}
		{\bibfnamefont {M.}~\bibnamefont {Segev}}, \ and\ \bibinfo {author}
		{\bibfnamefont {E.}~\bibnamefont {Karimi}},\ }\href {\doibase
		10.1080/00107514.2017.1418046} {\bibfield  {journal} {\bibinfo  {journal}
			{Contemp. Phys.}\ }\textbf {\bibinfo {volume} {59}},\ \bibinfo {pages} {126}
		(\bibinfo {year} {2018})},\ \Eprint
	{http://arxiv.org/abs/https://doi.org/10.1080/00107514.2017.1418046}
	{https://doi.org/10.1080/00107514.2017.1418046} \BibitemShut {NoStop}%
	\bibitem [{\citenamefont {Ivanov}(2022)}]{Ivanov2022}%
	\BibitemOpen
	\bibfield  {author} {\bibinfo {author} {\bibfnamefont {I.~P.}\ \bibnamefont
			{Ivanov}},\ }\href {\doibase https://doi.org/10.1016/j.ppnp.2022.103987}
	{\bibfield  {journal} {\bibinfo  {journal} {Prog. Part. Nucl. Phys.}\ ,\
			\bibinfo {pages} {103987}} (\bibinfo {year} {2022})}\BibitemShut {NoStop}%
	\bibitem [{\citenamefont {Bliokh}\ \emph {et~al.}(2007)\citenamefont {Bliokh},
		\citenamefont {Bliokh}, \citenamefont {Savel'ev},\ and\ \citenamefont
		{Nori}}]{Bliokh2007}%
	\BibitemOpen
	\bibfield  {author} {\bibinfo {author} {\bibfnamefont {K.~Y.}\ \bibnamefont
			{Bliokh}}, \bibinfo {author} {\bibfnamefont {Y.~P.}\ \bibnamefont {Bliokh}},
		\bibinfo {author} {\bibfnamefont {S.}~\bibnamefont {Savel'ev}}, \ and\
		\bibinfo {author} {\bibfnamefont {F.}~\bibnamefont {Nori}},\ }\href {\doibase
		10.1103/PhysRevLett.99.190404} {\bibfield  {journal} {\bibinfo  {journal}
			{Phys. Rev. Lett.}\ }\textbf {\bibinfo {volume} {99}},\ \bibinfo {pages}
		{190404} (\bibinfo {year} {2007})}\BibitemShut {NoStop}%
	\bibitem [{\citenamefont {Uchida}\ and\ \citenamefont
		{Tonomura}(2010)}]{Uchida2010}%
	\BibitemOpen
	\bibfield  {author} {\bibinfo {author} {\bibfnamefont {M.}~\bibnamefont
			{Uchida}}\ and\ \bibinfo {author} {\bibfnamefont {A.}~\bibnamefont
			{Tonomura}},\ }\href {\doibase 10.1038/nature08904} {\bibfield  {journal}
		{\bibinfo  {journal} {Nature}\ }\textbf {\bibinfo {volume} {464}},\ \bibinfo
		{pages} {737} (\bibinfo {year} {2010})}\BibitemShut {NoStop}%
	\bibitem [{\citenamefont {Verbeeck}, \citenamefont {Tian},\ and\ \citenamefont
		{Schattschneider}(2010)}]{Verbeeck2010}%
	\BibitemOpen
	\bibfield  {author} {\bibinfo {author} {\bibfnamefont {J.}~\bibnamefont
			{Verbeeck}}, \bibinfo {author} {\bibfnamefont {H.}~\bibnamefont {Tian}}, \
		and\ \bibinfo {author} {\bibfnamefont {P.}~\bibnamefont {Schattschneider}},\
	}\href {\doibase 10.1038/nature09366} {\bibfield  {journal} {\bibinfo
			{journal} {Nature}\ }\textbf {\bibinfo {volume} {467}},\ \bibinfo {pages}
		{301} (\bibinfo {year} {2010})}\BibitemShut {NoStop}%
	\bibitem [{\citenamefont {Mafakheri}\ \emph {et~al.}(2017)\citenamefont
		{Mafakheri}, \citenamefont {Tavabi}, \citenamefont {Lu}, \citenamefont
		{Balboni}, \citenamefont {Venturi}, \citenamefont {Menozzi}, \citenamefont
		{Gazzadi}, \citenamefont {Frabboni}, \citenamefont {Sit}, \citenamefont
		{Dunin-Borkowski}, \citenamefont {Karimi},\ and\ \citenamefont
		{Grillo}}]{Mafakheri2017}%
	\BibitemOpen
	\bibfield  {author} {\bibinfo {author} {\bibfnamefont {E.}~\bibnamefont
			{Mafakheri}}, \bibinfo {author} {\bibfnamefont {A.~H.}\ \bibnamefont
			{Tavabi}}, \bibinfo {author} {\bibfnamefont {P.-H.}\ \bibnamefont {Lu}},
		\bibinfo {author} {\bibfnamefont {R.}~\bibnamefont {Balboni}}, \bibinfo
		{author} {\bibfnamefont {F.}~\bibnamefont {Venturi}}, \bibinfo {author}
		{\bibfnamefont {C.}~\bibnamefont {Menozzi}}, \bibinfo {author} {\bibfnamefont
			{G.~C.}\ \bibnamefont {Gazzadi}}, \bibinfo {author} {\bibfnamefont
			{S.}~\bibnamefont {Frabboni}}, \bibinfo {author} {\bibfnamefont
			{A.}~\bibnamefont {Sit}}, \bibinfo {author} {\bibfnamefont {R.~E.}\
			\bibnamefont {Dunin-Borkowski}}, \bibinfo {author} {\bibfnamefont
			{E.}~\bibnamefont {Karimi}}, \ and\ \bibinfo {author} {\bibfnamefont
			{V.}~\bibnamefont {Grillo}},\ }\href {\doibase 10.1063/1.4977879} {\bibfield
		{journal} {\bibinfo  {journal} {Appl. Phys. Lett.}\ }\textbf {\bibinfo
			{volume} {110}},\ \bibinfo {pages} {093113} (\bibinfo {year}
		{2017})}\BibitemShut {NoStop}%
	\bibitem [{\citenamefont {Tavabi}\ \emph {et~al.}(2022)\citenamefont {Tavabi},
		\citenamefont {Rosi}, \citenamefont {Roncaglia}, \citenamefont {Rotunno},
		\citenamefont {Beleggia}, \citenamefont {Lu}, \citenamefont {Belsito},
		\citenamefont {Pozzi}, \citenamefont {Frabboni}, \citenamefont {Tiemeijer},
		\citenamefont {Dunin-Borkowski},\ and\ \citenamefont {Grillo}}]{Tavabi2022}%
	\BibitemOpen
	\bibfield  {author} {\bibinfo {author} {\bibfnamefont {A.~H.}\ \bibnamefont
			{Tavabi}}, \bibinfo {author} {\bibfnamefont {P.}~\bibnamefont {Rosi}},
		\bibinfo {author} {\bibfnamefont {A.}~\bibnamefont {Roncaglia}}, \bibinfo
		{author} {\bibfnamefont {E.}~\bibnamefont {Rotunno}}, \bibinfo {author}
		{\bibfnamefont {M.}~\bibnamefont {Beleggia}}, \bibinfo {author}
		{\bibfnamefont {P.-H.}\ \bibnamefont {Lu}}, \bibinfo {author} {\bibfnamefont
			{L.}~\bibnamefont {Belsito}}, \bibinfo {author} {\bibfnamefont
			{G.}~\bibnamefont {Pozzi}}, \bibinfo {author} {\bibfnamefont
			{S.}~\bibnamefont {Frabboni}}, \bibinfo {author} {\bibfnamefont
			{P.}~\bibnamefont {Tiemeijer}}, \bibinfo {author} {\bibfnamefont {R.~E.}\
			\bibnamefont {Dunin-Borkowski}}, \ and\ \bibinfo {author} {\bibfnamefont
			{V.}~\bibnamefont {Grillo}},\ }\href {\doibase 10.1063/5.0093411} {\bibfield
		{journal} {\bibinfo  {journal} {Applied Physics Letters}\ }\textbf {\bibinfo
			{volume} {121}},\ \bibinfo {pages} {073506} (\bibinfo {year}
		{2022})}\BibitemShut {NoStop}%
	\bibitem [{\citenamefont {Juchtmans}\ \emph {et~al.}(2015)\citenamefont
		{Juchtmans}, \citenamefont {B\'ech\'e}, \citenamefont {Abakumov},
		\citenamefont {Batuk},\ and\ \citenamefont {Verbeeck}}]{Juch2015}%
	\BibitemOpen
	\bibfield  {author} {\bibinfo {author} {\bibfnamefont {R.}~\bibnamefont
			{Juchtmans}}, \bibinfo {author} {\bibfnamefont {A.}~\bibnamefont
			{B\'ech\'e}}, \bibinfo {author} {\bibfnamefont {A.}~\bibnamefont {Abakumov}},
		\bibinfo {author} {\bibfnamefont {M.}~\bibnamefont {Batuk}}, \ and\ \bibinfo
		{author} {\bibfnamefont {J.}~\bibnamefont {Verbeeck}},\ }\href {\doibase
		10.1103/PhysRevB.91.094112} {\bibfield  {journal} {\bibinfo  {journal} {Phys.
				Rev. B}\ }\textbf {\bibinfo {volume} {91}},\ \bibinfo {pages} {094112}
		(\bibinfo {year} {2015})}\BibitemShut {NoStop}%
	\bibitem [{\citenamefont {Juchtmans}\ and\ \citenamefont
		{Verbeeck}(2016)}]{Juch2016b}%
	\BibitemOpen
	\bibfield  {author} {\bibinfo {author} {\bibfnamefont {R.}~\bibnamefont
			{Juchtmans}}\ and\ \bibinfo {author} {\bibfnamefont {J.}~\bibnamefont
			{Verbeeck}},\ }\href {\doibase 10.1103/PhysRevA.93.023811} {\bibfield
		{journal} {\bibinfo  {journal} {Phys. Rev. A}\ }\textbf {\bibinfo {volume}
			{93}},\ \bibinfo {pages} {023811} (\bibinfo {year} {2016})}\BibitemShut
	{NoStop}%
	\bibitem [{\citenamefont {Thirunavukkarasu}\ and\ \citenamefont {{G
				Thirunavukkarasu, J Yuan and M Babiker}}(2012)}]{Gnanavel2012}%
	\BibitemOpen
	\bibfield  {author} {\bibinfo {author} {\bibfnamefont {G.}~\bibnamefont
			{Thirunavukkarasu}}\ and\ \bibinfo {author} {\bibnamefont {{G
					Thirunavukkarasu, J Yuan and M Babiker}}}\ }(\bibinfo {year}
	{2012})\BibitemShut {NoStop}%
	\bibitem [{\citenamefont {Jesacher}\ \emph {et~al.}(2005)\citenamefont
		{Jesacher}, \citenamefont {F\"urhapter}, \citenamefont {Bernet},\ and\
		\citenamefont {Ritsch-Marte}}]{Jesacher2005}%
	\BibitemOpen
	\bibfield  {author} {\bibinfo {author} {\bibfnamefont {A.}~\bibnamefont
			{Jesacher}}, \bibinfo {author} {\bibfnamefont {S.}~\bibnamefont
			{F\"urhapter}}, \bibinfo {author} {\bibfnamefont {S.}~\bibnamefont {Bernet}},
		\ and\ \bibinfo {author} {\bibfnamefont {M.}~\bibnamefont {Ritsch-Marte}},\
	}\href {\doibase 10.1103/PhysRevLett.94.233902} {\bibfield  {journal}
		{\bibinfo  {journal} {Phys. Rev. Lett.}\ }\textbf {\bibinfo {volume} {94}},\
		\bibinfo {pages} {233902} (\bibinfo {year} {2005})}\BibitemShut {NoStop}%
	\bibitem [{\citenamefont {Serbo}\ \emph {et~al.}(2015)\citenamefont {Serbo},
		\citenamefont {Ivanov}, \citenamefont {Fritzsche}, \citenamefont {Seipt},\
		and\ \citenamefont {Surzhykov}}]{Serbo2015}%
	\BibitemOpen
	\bibfield  {author} {\bibinfo {author} {\bibfnamefont {V.}~\bibnamefont
			{Serbo}}, \bibinfo {author} {\bibfnamefont {I.~P.}\ \bibnamefont {Ivanov}},
		\bibinfo {author} {\bibfnamefont {S.}~\bibnamefont {Fritzsche}}, \bibinfo
		{author} {\bibfnamefont {D.}~\bibnamefont {Seipt}}, \ and\ \bibinfo {author}
		{\bibfnamefont {A.}~\bibnamefont {Surzhykov}},\ }\href {\doibase
		10.1103/PhysRevA.92.012705} {\bibfield  {journal} {\bibinfo  {journal} {Phys.
				Rev. A}\ }\textbf {\bibinfo {volume} {92}},\ \bibinfo {pages} {012705}
		(\bibinfo {year} {2015})}\BibitemShut {NoStop}%
	\bibitem [{\citenamefont {Dhankhar}\ and\ \citenamefont
		{Choubisa}(2020)}]{Dhankhar2020}%
	\BibitemOpen
	\bibfield  {author} {\bibinfo {author} {\bibfnamefont {N.}~\bibnamefont
			{Dhankhar}}\ and\ \bibinfo {author} {\bibfnamefont {R.}~\bibnamefont
			{Choubisa}},\ }\href {\doibase 10.1088/1361-6455/abcb52} {\bibfield
		{journal} {\bibinfo  {journal} {J. Phys. B: At. Mol. Opt. Phys.}\ }\textbf
		{\bibinfo {volume} {54}},\ \bibinfo {pages} {015203} (\bibinfo {year}
		{2020})}\BibitemShut {NoStop}%
	\bibitem [{\citenamefont {Gong}\ \emph {et~al.}(2022)\citenamefont {Gong},
		\citenamefont {Cheng}, \citenamefont {Zhang},\ and\ \citenamefont
		{Chen}}]{Gong2022}%
	\BibitemOpen
	\bibfield  {author} {\bibinfo {author} {\bibfnamefont {M.}~\bibnamefont
			{Gong}}, \bibinfo {author} {\bibfnamefont {Y.}~\bibnamefont {Cheng}},
		\bibinfo {author} {\bibfnamefont {S.~B.}\ \bibnamefont {Zhang}}, \ and\
		\bibinfo {author} {\bibfnamefont {X.}~\bibnamefont {Chen}},\ }\href {\doibase
		10.1103/PhysRevA.106.012818} {\bibfield  {journal} {\bibinfo  {journal}
			{Phys. Rev. A}\ }\textbf {\bibinfo {volume} {106}},\ \bibinfo {pages}
		{012818} (\bibinfo {year} {2022})}\BibitemShut {NoStop}%
	\bibitem [{\citenamefont {Bartschat}\ and\ \citenamefont
		{Kushner}(2016)}]{Bartschat2016}%
	\BibitemOpen
	\bibfield  {author} {\bibinfo {author} {\bibfnamefont {K.}~\bibnamefont
			{Bartschat}}\ and\ \bibinfo {author} {\bibfnamefont {M.~J.}\ \bibnamefont
			{Kushner}},\ }\href {\doibase 10.1073/pnas.1606132113} {\bibfield  {journal}
		{\bibinfo  {journal} {Proceedings of the National Academy of Sciences}\
		}\textbf {\bibinfo {volume} {113}},\ \bibinfo {pages} {7026} (\bibinfo {year}
		{2016})}\BibitemShut {NoStop}%
	\bibitem [{\citenamefont {Dunn}(2011)}]{DUNN2015}%
	\BibitemOpen
	\bibfield  {author} {\bibinfo {author} {\bibfnamefont {W.~B.}\ \bibnamefont
			{Dunn}},\ }in\ \href {\doibase
		https://doi.org/10.1016/B978-0-12-385118-5.00002-5} {\emph {\bibinfo
			{booktitle} {Methods in Systems Biology}}},\ \bibinfo {series} {Methods in
		Enzymology}, Vol.\ \bibinfo {volume} {500},\ \bibinfo {editor} {edited by\
		\bibinfo {editor} {\bibfnamefont {D.}~\bibnamefont {Jameson}}, \bibinfo
		{editor} {\bibfnamefont {M.}~\bibnamefont {Verma}}, \ and\ \bibinfo {editor}
		{\bibfnamefont {H.~V.}\ \bibnamefont {Westerhoff}}}\ (\bibinfo  {publisher}
	{Academic Press},\ \bibinfo {year} {2011})\ pp.\ \bibinfo {pages}
	{15--35}\BibitemShut {NoStop}%
	\bibitem [{\citenamefont {Girazian}\ \emph {et~al.}(2017)\citenamefont
		{Girazian}, \citenamefont {Mahaffy}, \citenamefont {Lillis}, \citenamefont
		{Benna}, \citenamefont {Elrod}, \citenamefont {Fowler},\ and\ \citenamefont
		{Mitchell}}]{Girazian2017}%
	\BibitemOpen
	\bibfield  {author} {\bibinfo {author} {\bibfnamefont {Z.}~\bibnamefont
			{Girazian}}, \bibinfo {author} {\bibfnamefont {P.}~\bibnamefont {Mahaffy}},
		\bibinfo {author} {\bibfnamefont {R.~J.}\ \bibnamefont {Lillis}}, \bibinfo
		{author} {\bibfnamefont {M.}~\bibnamefont {Benna}}, \bibinfo {author}
		{\bibfnamefont {M.}~\bibnamefont {Elrod}}, \bibinfo {author} {\bibfnamefont
			{C.~M.}\ \bibnamefont {Fowler}}, \ and\ \bibinfo {author} {\bibfnamefont
			{D.~L.}\ \bibnamefont {Mitchell}},\ }\href {\doibase
		https://doi.org/10.1002/2017GL075431} {\bibfield  {journal} {\bibinfo
			{journal} {Geophys. Res. Lett.}\ }\textbf {\bibinfo {volume} {44}},\ \bibinfo
		{pages} {11,248} (\bibinfo {year} {2017})}\BibitemShut {NoStop}%
	\bibitem [{\citenamefont {{Kyniene, Ausra}}\ \emph {et~al.}(2019)\citenamefont
		{{Kyniene, Ausra}}, \citenamefont {{Kucas, Sigitas}}, \citenamefont {{Masys,
				Sarunas}},\ and\ \citenamefont {{Jonauskas, Valdas}}}]{Kyniene2019}%
	\BibitemOpen
	\bibfield  {author} {\bibinfo {author} {\bibnamefont {{Kyniene, Ausra}}},
		\bibinfo {author} {\bibnamefont {{Kucas, Sigitas}}}, \bibinfo {author}
		{\bibnamefont {{Masys, Sarunas}}}, \ and\ \bibinfo {author} {\bibnamefont
			{{Jonauskas, Valdas}}},\ }\href {\doibase 10.1051/0004-6361/201833762}
	{\bibfield  {journal} {\bibinfo  {journal} {A\&A}\ }\textbf {\bibinfo
			{volume} {624}},\ \bibinfo {pages} {A14} (\bibinfo {year}
		{2019})}\BibitemShut {NoStop}%
	\bibitem [{\citenamefont {Caleman}\ \emph {et~al.}(2009)\citenamefont
		{Caleman}, \citenamefont {Ortiz}, \citenamefont {Marklund}, \citenamefont
		{Bultmark}, \citenamefont {Gabrysch}, \citenamefont {Parak}, \citenamefont
		{Hajdu}, \citenamefont {Klintenberg},\ and\ \citenamefont
		{T{\^{\i}}mneanu}}]{Caleman2009}%
	\BibitemOpen
	\bibfield  {author} {\bibinfo {author} {\bibfnamefont {C.}~\bibnamefont
			{Caleman}}, \bibinfo {author} {\bibfnamefont {C.}~\bibnamefont {Ortiz}},
		\bibinfo {author} {\bibfnamefont {E.}~\bibnamefont {Marklund}}, \bibinfo
		{author} {\bibfnamefont {F.}~\bibnamefont {Bultmark}}, \bibinfo {author}
		{\bibfnamefont {M.}~\bibnamefont {Gabrysch}}, \bibinfo {author}
		{\bibfnamefont {F.~G.}\ \bibnamefont {Parak}}, \bibinfo {author}
		{\bibfnamefont {J.}~\bibnamefont {Hajdu}}, \bibinfo {author} {\bibfnamefont
			{M.}~\bibnamefont {Klintenberg}}, \ and\ \bibinfo {author} {\bibfnamefont
			{N.}~\bibnamefont {T{\^{\i}}mneanu}},\ }\href {\doibase
		10.1209/0295-5075/85/18005} {\bibfield  {journal} {\bibinfo  {journal} {EPL
				(Europhysics Letters)}\ }\textbf {\bibinfo {volume} {85}},\ \bibinfo {pages}
		{18005} (\bibinfo {year} {2009})}\BibitemShut {NoStop}%
	\bibitem [{\citenamefont {Yavuz}\ \emph {et~al.}(2014)\citenamefont {Yavuz},
		\citenamefont {Okumus}, \citenamefont {Ozer}, \citenamefont {Ulu},
		\citenamefont {Dogan}, \citenamefont {Sahlaoui}, \citenamefont {Benmansour},\
		and\ \citenamefont {Bouamoud}}]{Yavuz2014}%
	\BibitemOpen
	\bibfield  {author} {\bibinfo {author} {\bibfnamefont {M.}~\bibnamefont
			{Yavuz}}, \bibinfo {author} {\bibfnamefont {N.}~\bibnamefont {Okumus}},
		\bibinfo {author} {\bibfnamefont {Z.~N.}\ \bibnamefont {Ozer}}, \bibinfo
		{author} {\bibfnamefont {M.}~\bibnamefont {Ulu}}, \bibinfo {author}
		{\bibfnamefont {M.}~\bibnamefont {Dogan}}, \bibinfo {author} {\bibfnamefont
			{M.}~\bibnamefont {Sahlaoui}}, \bibinfo {author} {\bibfnamefont
			{H.}~\bibnamefont {Benmansour}}, \ and\ \bibinfo {author} {\bibfnamefont
			{M.}~\bibnamefont {Bouamoud}},\ }\href {\doibase
		10.1088/1742-6596/488/5/052031} {\bibfield  {journal} {\bibinfo  {journal}
			{J. Phys: Conference Series}\ }\textbf {\bibinfo {volume} {488}},\ \bibinfo
		{pages} {052031} (\bibinfo {year} {2014})}\BibitemShut {NoStop}%
	\bibitem [{\citenamefont {Yavuz}\ \emph {et~al.}(2016)\citenamefont {Yavuz},
		\citenamefont {Ozer}, \citenamefont {Ulu}, \citenamefont {Champion},\ and\
		\citenamefont {Dogan}}]{Yavuz2016}%
	\BibitemOpen
	\bibfield  {author} {\bibinfo {author} {\bibfnamefont {M.}~\bibnamefont
			{Yavuz}}, \bibinfo {author} {\bibfnamefont {Z.~N.}\ \bibnamefont {Ozer}},
		\bibinfo {author} {\bibfnamefont {M.}~\bibnamefont {Ulu}}, \bibinfo {author}
		{\bibfnamefont {C.}~\bibnamefont {Champion}}, \ and\ \bibinfo {author}
		{\bibfnamefont {M.}~\bibnamefont {Dogan}},\ }\href {\doibase
		10.1063/1.4947591} {\bibfield  {journal} {\bibinfo  {journal} {J. Chem.
				Phys.}\ }\textbf {\bibinfo {volume} {144}},\ \bibinfo {pages} {164305}
		(\bibinfo {year} {2016})},\ \Eprint
	{http://arxiv.org/abs/https://doi.org/10.1063/1.4947591}
	{https://doi.org/10.1063/1.4947591} \BibitemShut {NoStop}%
	\bibitem [{\citenamefont {Tachino}\ \emph {et~al.}(2015)\citenamefont
		{Tachino}, \citenamefont {Monti}, \citenamefont {Fojón}, \citenamefont
		{Champion},\ and\ \citenamefont {Rivarola}}]{Tachino2015}%
	\BibitemOpen
	\bibfield  {author} {\bibinfo {author} {\bibfnamefont {C.~A.}\ \bibnamefont
			{Tachino}}, \bibinfo {author} {\bibfnamefont {J.~M.}\ \bibnamefont {Monti}},
		\bibinfo {author} {\bibfnamefont {O.~A.}\ \bibnamefont {Fojón}}, \bibinfo
		{author} {\bibfnamefont {C.}~\bibnamefont {Champion}}, \ and\ \bibinfo
		{author} {\bibfnamefont {R.~D.}\ \bibnamefont {Rivarola}},\ }\href {\doibase
		10.1088/1742-6596/583/1/012020} {\bibfield  {journal} {\bibinfo  {journal}
			{Journal of Physics: Conference Series}\ }\textbf {\bibinfo {volume} {583}},\
		\bibinfo {pages} {012020} (\bibinfo {year} {2015})}\BibitemShut {NoStop}%
	\bibitem [{\citenamefont {T{\'o}th}, \citenamefont {Nagy},\ and\ \citenamefont
		{Campeanu}(2016)}]{Toth2016}%
	\BibitemOpen
	\bibfield  {author} {\bibinfo {author} {\bibfnamefont {I.}~\bibnamefont
			{T{\'o}th}}, \bibinfo {author} {\bibfnamefont {L.}~\bibnamefont {Nagy}}, \
		and\ \bibinfo {author} {\bibfnamefont {R.~I.}\ \bibnamefont {Campeanu}},\
	}\href {\doibase 10.1140/epjd/e2016-70135-4} {\bibfield  {journal} {\bibinfo
			{journal} {The European Physical Journal D}\ }\textbf {\bibinfo {volume}
			{70}},\ \bibinfo {pages} {170} (\bibinfo {year} {2016})}\BibitemShut
	{NoStop}%
	\bibitem [{\citenamefont {Lahmam-Bennani}\ \emph {et~al.}(2009)\citenamefont
		{Lahmam-Bennani}, \citenamefont {Naja}, \citenamefont {Casagrande},
		\citenamefont {Okumus}, \citenamefont {Cappello}, \citenamefont
		{Charpentier},\ and\ \citenamefont {Houamer}}]{LB2009}%
	\BibitemOpen
	\bibfield  {author} {\bibinfo {author} {\bibfnamefont {A.}~\bibnamefont
			{Lahmam-Bennani}}, \bibinfo {author} {\bibfnamefont {A.}~\bibnamefont
			{Naja}}, \bibinfo {author} {\bibfnamefont {E.~M.~S.}\ \bibnamefont
			{Casagrande}}, \bibinfo {author} {\bibfnamefont {N.}~\bibnamefont {Okumus}},
		\bibinfo {author} {\bibfnamefont {C.~D.}\ \bibnamefont {Cappello}}, \bibinfo
		{author} {\bibfnamefont {I.}~\bibnamefont {Charpentier}}, \ and\ \bibinfo
		{author} {\bibfnamefont {S.}~\bibnamefont {Houamer}},\ }\href {\doibase
		10.1088/0953-4075/42/16/165201} {\bibfield  {journal} {\bibinfo  {journal}
			{J. Phys. B: At. Mol. Opt. Phys.: At. Mol. Opt. Phys.}\ }\textbf {\bibinfo
			{volume} {42}},\ \bibinfo {pages} {165201} (\bibinfo {year}
		{2009})}\BibitemShut {NoStop}%
	\bibitem [{\citenamefont {Mir}\ \emph {et~al.}(2015)\citenamefont {Mir},
		\citenamefont {Casagrande}, \citenamefont {Naja}, \citenamefont {Cappello},
		\citenamefont {Houamer},\ and\ \citenamefont {Omar}}]{Mir2015}%
	\BibitemOpen
	\bibfield  {author} {\bibinfo {author} {\bibfnamefont {R.~E.}\ \bibnamefont
			{Mir}}, \bibinfo {author} {\bibfnamefont {E.~M.~S.}\ \bibnamefont
			{Casagrande}}, \bibinfo {author} {\bibfnamefont {A.}~\bibnamefont {Naja}},
		\bibinfo {author} {\bibfnamefont {C.~D.}\ \bibnamefont {Cappello}}, \bibinfo
		{author} {\bibfnamefont {S.}~\bibnamefont {Houamer}}, \ and\ \bibinfo
		{author} {\bibfnamefont {F.~E.}\ \bibnamefont {Omar}},\ }\href {\doibase
		10.1088/0953-4075/48/17/175202} {\bibfield  {journal} {\bibinfo  {journal}
			{J. Phys. B: At. Mol. Opt. Phys.}\ }\textbf {\bibinfo {volume} {48}},\
		\bibinfo {pages} {175202} (\bibinfo {year} {2015})}\BibitemShut {NoStop}%
	\bibitem [{\citenamefont {{Mouawad, Lena}}\ \emph {et~al.}(2018)\citenamefont
		{{Mouawad, Lena}}, \citenamefont {{Bitar, Ziad El}}, \citenamefont {{Osman,
				Ahmad}}, \citenamefont {{Khalil, Mohamad}}, \citenamefont {{Hervieux, Paul
				Antoine}},\ and\ \citenamefont {{Cappello, Claude Dal}}}]{Mouawad2018}%
	\BibitemOpen
	\bibfield  {author} {\bibinfo {author} {\bibnamefont {{Mouawad, Lena}}},
		\bibinfo {author} {\bibnamefont {{Bitar, Ziad El}}}, \bibinfo {author}
		{\bibnamefont {{Osman, Ahmad}}}, \bibinfo {author} {\bibnamefont {{Khalil,
					Mohamad}}}, \bibinfo {author} {\bibnamefont {{Hervieux, Paul Antoine}}}, \
		and\ \bibinfo {author} {\bibnamefont {{Cappello, Claude Dal}}},\ }\href
	{\doibase 10.1051/epjconf/201817001012} {\bibfield  {journal} {\bibinfo
			{journal} {EPJ Web Conf.}\ }\textbf {\bibinfo {volume} {170}},\ \bibinfo
		{pages} {01012} (\bibinfo {year} {2018})}\BibitemShut {NoStop}%
	\bibitem [{\citenamefont {Bouchikhi}\ \emph {et~al.}(2018)\citenamefont
		{Bouchikhi}, \citenamefont {Sahlaoui}, \citenamefont {Lasri}, \citenamefont
		{Sekkal},\ and\ \citenamefont {Bouamoud}}]{Bouchikhi2019}%
	\BibitemOpen
	\bibfield  {author} {\bibinfo {author} {\bibfnamefont {A.}~\bibnamefont
			{Bouchikhi}}, \bibinfo {author} {\bibfnamefont {M.}~\bibnamefont {Sahlaoui}},
		\bibinfo {author} {\bibfnamefont {B.}~\bibnamefont {Lasri}}, \bibinfo
		{author} {\bibfnamefont {A.}~\bibnamefont {Sekkal}}, \ and\ \bibinfo {author}
		{\bibfnamefont {M.}~\bibnamefont {Bouamoud}},\ }\href {\doibase
		10.1088/1361-6455/aada77} {\bibfield  {journal} {\bibinfo  {journal} {J.
				Phys. B: At. Mol. Opt. Phys.}\ }\textbf {\bibinfo {volume} {52}},\ \bibinfo
		{pages} {015201} (\bibinfo {year} {2018})}\BibitemShut {NoStop}%
	\bibitem [{\citenamefont {Nixon}\ \emph {et~al.}(2011)\citenamefont {Nixon},
		\citenamefont {Murray}, \citenamefont {Chaluvadi}, \citenamefont {Ning},\
		and\ \citenamefont {Madison}}]{Cnixon2011}%
	\BibitemOpen
	\bibfield  {author} {\bibinfo {author} {\bibfnamefont {K.~L.}\ \bibnamefont
			{Nixon}}, \bibinfo {author} {\bibfnamefont {A.~J.}\ \bibnamefont {Murray}},
		\bibinfo {author} {\bibfnamefont {H.}~\bibnamefont {Chaluvadi}}, \bibinfo
		{author} {\bibfnamefont {C.}~\bibnamefont {Ning}}, \ and\ \bibinfo {author}
		{\bibfnamefont {D.~H.}\ \bibnamefont {Madison}},\ }\href {\doibase
		10.1063/1.3581812} {\bibfield  {journal} {\bibinfo  {journal} {J. Chem.
				Phys.}\ }\textbf {\bibinfo {volume} {134}},\ \bibinfo {pages} {174304}
		(\bibinfo {year} {2011})},\ \Eprint
	{http://arxiv.org/abs/https://doi.org/10.1063/1.3581812}
	{https://doi.org/10.1063/1.3581812} \BibitemShut {NoStop}%
	\bibitem [{\citenamefont {Ali}\ \emph {et~al.}(2019)\citenamefont {Ali},
		\citenamefont {Granados}, \citenamefont {Sakaamini}, \citenamefont {Harvey},
		\citenamefont {Ancarani}, \citenamefont {Murray}, \citenamefont {Dogan},
		\citenamefont {Ning}, \citenamefont {Colgan},\ and\ \citenamefont
		{Madison}}]{Ali2019}%
	\BibitemOpen
	\bibfield  {author} {\bibinfo {author} {\bibfnamefont {E.}~\bibnamefont
			{Ali}}, \bibinfo {author} {\bibfnamefont {C.}~\bibnamefont {Granados}},
		\bibinfo {author} {\bibfnamefont {A.}~\bibnamefont {Sakaamini}}, \bibinfo
		{author} {\bibfnamefont {M.}~\bibnamefont {Harvey}}, \bibinfo {author}
		{\bibfnamefont {L.~U.}\ \bibnamefont {Ancarani}}, \bibinfo {author}
		{\bibfnamefont {A.~J.}\ \bibnamefont {Murray}}, \bibinfo {author}
		{\bibfnamefont {M.}~\bibnamefont {Dogan}}, \bibinfo {author} {\bibfnamefont
			{C.}~\bibnamefont {Ning}}, \bibinfo {author} {\bibfnamefont {J.}~\bibnamefont
			{Colgan}}, \ and\ \bibinfo {author} {\bibfnamefont {D.}~\bibnamefont
			{Madison}},\ }\href {\doibase 10.1063/1.5097670} {\bibfield  {journal}
		{\bibinfo  {journal} {J. Chem. Phys.}\ }\textbf {\bibinfo {volume} {150}},\
		\bibinfo {pages} {194302} (\bibinfo {year} {2019})},\ \Eprint
	{http://arxiv.org/abs/https://doi.org/10.1063/1.5097670}
	{https://doi.org/10.1063/1.5097670} \BibitemShut {NoStop}%
	\bibitem [{\citenamefont {Nixon}\ \emph {et~al.}(2013)\citenamefont {Nixon},
		\citenamefont {Murray}, \citenamefont {Chaluvadi}, \citenamefont {Ning},
		\citenamefont {Colgan},\ and\ \citenamefont {Madison}}]{Nixon2013}%
	\BibitemOpen
	\bibfield  {author} {\bibinfo {author} {\bibfnamefont {K.~L.}\ \bibnamefont
			{Nixon}}, \bibinfo {author} {\bibfnamefont {A.~J.}\ \bibnamefont {Murray}},
		\bibinfo {author} {\bibfnamefont {H.}~\bibnamefont {Chaluvadi}}, \bibinfo
		{author} {\bibfnamefont {C.}~\bibnamefont {Ning}}, \bibinfo {author}
		{\bibfnamefont {J.}~\bibnamefont {Colgan}}, \ and\ \bibinfo {author}
		{\bibfnamefont {D.~H.}\ \bibnamefont {Madison}},\ }\href {\doibase
		10.1063/1.4802960} {\bibfield  {journal} {\bibinfo  {journal} {J. Chem.
				Phys.}\ }\textbf {\bibinfo {volume} {138}},\ \bibinfo {pages} {174304}
		(\bibinfo {year} {2013})}\BibitemShut {NoStop}%
	\bibitem [{\citenamefont {Tóth}\ and\ \citenamefont {Nagy}(2010)}]{Toth2010}%
	\BibitemOpen
	\bibfield  {author} {\bibinfo {author} {\bibfnamefont {I.}~\bibnamefont
			{Tóth}}\ and\ \bibinfo {author} {\bibfnamefont {L.}~\bibnamefont {Nagy}},\
	}\href {\doibase 10.1088/0953-4075/43/13/135204} {\bibfield  {journal}
		{\bibinfo  {journal} {J. Phys. B: At. Mol. Opt. Phys.: At. Mol. Opt. Phys.}\
		}\textbf {\bibinfo {volume} {43}},\ \bibinfo {pages} {135204} (\bibinfo
		{year} {2010})}\BibitemShut {NoStop}%
	\bibitem [{\citenamefont {Lin}, \citenamefont {McCurdy},\ and\ \citenamefont
		{Rescigno}(2014)}]{Chih2014}%
	\BibitemOpen
	\bibfield  {author} {\bibinfo {author} {\bibfnamefont {C.-Y.}\ \bibnamefont
			{Lin}}, \bibinfo {author} {\bibfnamefont {C.~W.}\ \bibnamefont {McCurdy}}, \
		and\ \bibinfo {author} {\bibfnamefont {T.~N.}\ \bibnamefont {Rescigno}},\
	}\href {\doibase 10.1103/PhysRevA.89.052718} {\bibfield  {journal} {\bibinfo
			{journal} {Phys. Rev. A}\ }\textbf {\bibinfo {volume} {89}},\ \bibinfo
		{pages} {052718} (\bibinfo {year} {2014})}\BibitemShut {NoStop}%
	\bibitem [{\citenamefont {Castro}(2016)}]{Castro2016}%
	\BibitemOpen
	\bibfield  {author} {\bibinfo {author} {\bibfnamefont {C.~M.~G.}\
			\bibnamefont {Castro}},\ }\emph {\bibinfo {title} {Application of generalized
			Sturmian basis functions to molecular systems}},\ \href@noop {} {Ph.D.
		thesis},\ \bibinfo  {school} {Universit{\'e} de Lorraine} (\bibinfo {year}
	{2016})\BibitemShut {NoStop}%
	\bibitem [{\citenamefont {Granados-Castro}\ and\ \citenamefont
		{Ancarani}(2017)}]{Castro2017}%
	\BibitemOpen
	\bibfield  {author} {\bibinfo {author} {\bibfnamefont {C.~M.}\ \bibnamefont
			{Granados-Castro}}\ and\ \bibinfo {author} {\bibfnamefont {L.~U.}\
			\bibnamefont {Ancarani}},\ }\href {\doibase 10.1140/epjd/e2017-70721-x}
	{\bibfield  {journal} {\bibinfo  {journal} {The European Physical Journal D}\
		}\textbf {\bibinfo {volume} {71}},\ \bibinfo {pages} {65} (\bibinfo {year}
		{2017})}\BibitemShut {NoStop}%
	\bibitem [{\citenamefont {Gong}\ \emph {et~al.}(2017)\citenamefont {Gong},
		\citenamefont {Li}, \citenamefont {Zhang}, \citenamefont {Liu}, \citenamefont
		{Wu}, \citenamefont {Wang}, \citenamefont {Qu},\ and\ \citenamefont
		{Chen}}]{Gong2017}%
	\BibitemOpen
	\bibfield  {author} {\bibinfo {author} {\bibfnamefont {M.}~\bibnamefont
			{Gong}}, \bibinfo {author} {\bibfnamefont {X.}~\bibnamefont {Li}}, \bibinfo
		{author} {\bibfnamefont {S.~B.}\ \bibnamefont {Zhang}}, \bibinfo {author}
		{\bibfnamefont {L.}~\bibnamefont {Liu}}, \bibinfo {author} {\bibfnamefont
			{Y.}~\bibnamefont {Wu}}, \bibinfo {author} {\bibfnamefont {J.}~\bibnamefont
			{Wang}}, \bibinfo {author} {\bibfnamefont {Y.}~\bibnamefont {Qu}}, \ and\
		\bibinfo {author} {\bibfnamefont {X.}~\bibnamefont {Chen}},\ }\href {\doibase
		10.1103/PhysRevA.96.042703} {\bibfield  {journal} {\bibinfo  {journal} {Phys.
				Rev. A}\ }\textbf {\bibinfo {volume} {96}},\ \bibinfo {pages} {042703}
		(\bibinfo {year} {2017})}\BibitemShut {NoStop}%
	\bibitem [{\citenamefont {Houamer}, \citenamefont {Chinoune},\ and\
		\citenamefont {Cappello}(2017)}]{Houamer2017}%
	\BibitemOpen
	\bibfield  {author} {\bibinfo {author} {\bibfnamefont {S.}~\bibnamefont
			{Houamer}}, \bibinfo {author} {\bibfnamefont {M.}~\bibnamefont {Chinoune}}, \
		and\ \bibinfo {author} {\bibfnamefont {C.~D.}\ \bibnamefont {Cappello}},\
	}\href {\doibase 10.1140/epjd/e2016-70596-3} {\bibfield  {journal} {\bibinfo
			{journal} {The European Physical Journal D}\ }\textbf {\bibinfo {volume}
			{71}},\ \bibinfo {pages} {17} (\bibinfo {year} {2017})}\BibitemShut {NoStop}%
	\bibitem [{\citenamefont {Mir}\ \emph {et~al.}(2020)\citenamefont {Mir},
		\citenamefont {Kaja}, \citenamefont {Naja}, \citenamefont {Casagrande},
		\citenamefont {Houamer},\ and\ \citenamefont {Cappello}}]{Mir2020}%
	\BibitemOpen
	\bibfield  {author} {\bibinfo {author} {\bibfnamefont {R.~E.}\ \bibnamefont
			{Mir}}, \bibinfo {author} {\bibfnamefont {K.}~\bibnamefont {Kaja}}, \bibinfo
		{author} {\bibfnamefont {A.}~\bibnamefont {Naja}}, \bibinfo {author}
		{\bibfnamefont {E.~M.~S.}\ \bibnamefont {Casagrande}}, \bibinfo {author}
		{\bibfnamefont {S.}~\bibnamefont {Houamer}}, \ and\ \bibinfo {author}
		{\bibfnamefont {C.~D.}\ \bibnamefont {Cappello}},\ }\href {\doibase
		10.1088/1361-6455/abc144} {\bibfield  {journal} {\bibinfo  {journal} {J.
				Phys. B: At. Mol. Opt. Phys.}\ }\textbf {\bibinfo {volume} {54}},\ \bibinfo
		{pages} {015201} (\bibinfo {year} {2020})}\BibitemShut {NoStop}%
	\bibitem [{\citenamefont {Harris}, \citenamefont {Plumadore},\ and\
		\citenamefont {Smozhanyk}(2019)}]{Harris2019}%
	\BibitemOpen
	\bibfield  {author} {\bibinfo {author} {\bibfnamefont {A.~L.}\ \bibnamefont
			{Harris}}, \bibinfo {author} {\bibfnamefont {A.}~\bibnamefont {Plumadore}}, \
		and\ \bibinfo {author} {\bibfnamefont {Z.}~\bibnamefont {Smozhanyk}},\ }\href
	{\doibase 10.1088/1361-6455/ab12f3} {\bibfield  {journal} {\bibinfo
			{journal} {J. Phys. B: At. Mol. Opt. Phys.}\ }\textbf {\bibinfo {volume}
			{52}},\ \bibinfo {pages} {094001} (\bibinfo {year} {2019})}\BibitemShut
	{NoStop}%
	\bibitem [{\citenamefont {Plumadore}\ and\ \citenamefont
		{Harris}(2020)}]{Plumadore2020}%
	\BibitemOpen
	\bibfield  {author} {\bibinfo {author} {\bibfnamefont {A.}~\bibnamefont
			{Plumadore}}\ and\ \bibinfo {author} {\bibfnamefont {A.~L.}\ \bibnamefont
			{Harris}},\ }\href {\doibase 10.1088/1361-6455/abb3ac} {\bibfield  {journal}
		{\bibinfo  {journal} {J. Phys. B: At. Mol. Opt. Phys.}\ }\textbf {\bibinfo
			{volume} {53}},\ \bibinfo {pages} {205205} (\bibinfo {year}
		{2020})}\BibitemShut {NoStop}%
	\bibitem [{\citenamefont {Dhankhar}\ and\ \citenamefont
		{Choubisa}(2022)}]{Dhankhar2022}%
	\BibitemOpen
	\bibfield  {author} {\bibinfo {author} {\bibfnamefont {N.}~\bibnamefont
			{Dhankhar}}\ and\ \bibinfo {author} {\bibfnamefont {R.}~\bibnamefont
			{Choubisa}},\ }\href {\doibase 10.1103/PhysRevA.105.062801} {\bibfield
		{journal} {\bibinfo  {journal} {Phys. Rev. A}\ }\textbf {\bibinfo {volume}
			{105}},\ \bibinfo {pages} {062801} (\bibinfo {year} {2022})}\BibitemShut
	{NoStop}%
	\bibitem [{\citenamefont {Dhankhar}, \citenamefont {Banerjee},\ and\
		\citenamefont {Choubisa}(2022)}]{Dhankhar2022_2}%
	\BibitemOpen
	\bibfield  {author} {\bibinfo {author} {\bibfnamefont {N.}~\bibnamefont
			{Dhankhar}}, \bibinfo {author} {\bibfnamefont {S.}~\bibnamefont {Banerjee}},
		\ and\ \bibinfo {author} {\bibfnamefont {R.}~\bibnamefont {Choubisa}},\
	}\href {\doibase 10.1088/1361-6455/ac7d80} {\bibfield  {journal} {\bibinfo
			{journal} {J. Phys. B: At. Mol. Opt. Phys.: At. Mol. Opt. Phys.}\ }\textbf
		{\bibinfo {volume} {55}},\ \bibinfo {pages} {165202} (\bibinfo {year}
		{2022})}\BibitemShut {NoStop}%
	\bibitem [{\citenamefont {Mandal}\ \emph {et~al.}(2021)\citenamefont {Mandal},
		\citenamefont {Dhankhar}, \citenamefont {S\'ebilleau},\ and\ \citenamefont
		{Choubisa}}]{Mandal2021}%
	\BibitemOpen
	\bibfield  {author} {\bibinfo {author} {\bibfnamefont {A.}~\bibnamefont
			{Mandal}}, \bibinfo {author} {\bibfnamefont {N.}~\bibnamefont {Dhankhar}},
		\bibinfo {author} {\bibfnamefont {D.}~\bibnamefont {S\'ebilleau}}, \ and\
		\bibinfo {author} {\bibfnamefont {R.}~\bibnamefont {Choubisa}},\ }\href
	{\doibase 10.1103/PhysRevA.104.052818} {\bibfield  {journal} {\bibinfo
			{journal} {Phys. Rev. A}\ }\textbf {\bibinfo {volume} {104}},\ \bibinfo
		{pages} {052818} (\bibinfo {year} {2021})}\BibitemShut {NoStop}%
	\bibitem [{\citenamefont {Moccia}(1964{\natexlab{a}})}]{Moccia1}%
	\BibitemOpen
	\bibfield  {author} {\bibinfo {author} {\bibfnamefont {R.}~\bibnamefont
			{Moccia}},\ }\href {\doibase 10.1063/1.1725489} {\bibfield  {journal}
		{\bibinfo  {journal} {J. Chem. Phys.}\ }\textbf {\bibinfo {volume} {40}},\
		\bibinfo {pages} {2164} (\bibinfo {year} {1964}{\natexlab{a}})},\ \Eprint
	{http://arxiv.org/abs/https://doi.org/10.1063/1.1725489}
	{https://doi.org/10.1063/1.1725489} \BibitemShut {NoStop}%
	\bibitem [{\citenamefont {Moccia}(1964{\natexlab{b}})}]{Moccia2}%
	\BibitemOpen
	\bibfield  {author} {\bibinfo {author} {\bibfnamefont {R.}~\bibnamefont
			{Moccia}},\ }\href {\doibase 10.1063/1.1725490} {\bibfield  {journal}
		{\bibinfo  {journal} {J. Chem. Phys.}\ }\textbf {\bibinfo {volume} {40}},\
		\bibinfo {pages} {2176} (\bibinfo {year} {1964}{\natexlab{b}})},\ \Eprint
	{http://arxiv.org/abs/https://doi.org/10.1063/1.1725490}
	{https://doi.org/10.1063/1.1725490} \BibitemShut {NoStop}%
	\bibitem [{\citenamefont {Champion}, \citenamefont {Hanssen},\ and\
		\citenamefont {Hervieux}(2005)}]{Champion2005}%
	\BibitemOpen
	\bibfield  {author} {\bibinfo {author} {\bibfnamefont {C.}~\bibnamefont
			{Champion}}, \bibinfo {author} {\bibfnamefont {J.}~\bibnamefont {Hanssen}}, \
		and\ \bibinfo {author} {\bibfnamefont {P.~A.}\ \bibnamefont {Hervieux}},\
	}\href {\doibase 10.1103/PhysRevA.72.059906} {\bibfield  {journal} {\bibinfo
			{journal} {Phys. Rev. A}\ }\textbf {\bibinfo {volume} {72}},\ \bibinfo
		{pages} {059906} (\bibinfo {year} {2005})}\BibitemShut {NoStop}%
	\bibitem [{\citenamefont {Karlovets}\ \emph {et~al.}(2017)\citenamefont
		{Karlovets}, \citenamefont {Kotkin}, \citenamefont {Serbo},\ and\
		\citenamefont {Surzhykov}}]{Karlovets2017}%
	\BibitemOpen
	\bibfield  {author} {\bibinfo {author} {\bibfnamefont {D.~V.}\ \bibnamefont
			{Karlovets}}, \bibinfo {author} {\bibfnamefont {G.~L.}\ \bibnamefont
			{Kotkin}}, \bibinfo {author} {\bibfnamefont {V.~G.}\ \bibnamefont {Serbo}}, \
		and\ \bibinfo {author} {\bibfnamefont {A.}~\bibnamefont {Surzhykov}},\ }\href
	{\doibase 10.1103/PhysRevA.95.032703} {\bibfield  {journal} {\bibinfo
			{journal} {Phys. Rev. A}\ }\textbf {\bibinfo {volume} {95}},\ \bibinfo
		{pages} {032703} (\bibinfo {year} {2017})}\BibitemShut {NoStop}%
	\bibitem [{\citenamefont {Khajuria}\ and\ \citenamefont
		{Tripathi}(1999)}]{Khajuria1999}%
	\BibitemOpen
	\bibfield  {author} {\bibinfo {author} {\bibfnamefont {Y.}~\bibnamefont
			{Khajuria}}\ and\ \bibinfo {author} {\bibfnamefont {D.~N.}\ \bibnamefont
			{Tripathi}},\ }\href {\doibase 10.1103/PhysRevA.59.1197} {\bibfield
		{journal} {\bibinfo  {journal} {Phys. Rev. A}\ }\textbf {\bibinfo {volume}
			{59}},\ \bibinfo {pages} {1197} (\bibinfo {year} {1999})}\BibitemShut
	{NoStop}%
\end{thebibliography}

%
\end{document}